\documentstyle[aps,12pt]{revtex}

\begin{document}

%\begin{frontmatter}
\title{Deterministic Chaos and the Foundations of the Kinetic Theory of Gases}
\author{J. R. Dorfman}
\address{Institute for Physical Science and Technology and Department
of Physics, \\ University of Maryland, \\ College Park, Maryland,
20742, USA}
\date{\today}
\maketitle
\begin{abstract}
Recent work in dynamical systems theory has shown that many properties
that are associated with irreversible processes in fluids can be
understood in terms of the dynamical properties of reversible,
Hamiltonian systems. That is, stochastic-like behavior is possible for
these systems. Here we review the basic theory for
this stochastic-like behavior and show how it may be used to obtain an
understanding of irreversible processes in gases and fluids. Recent,
closely related,
work on the use of kinetic theory to calculate dynamical
quantities such as Lyapunov exponents is also discussed.
\end{abstract}
%\end{frontmatter}

\section{Introduction}

Dynamical systems theory \cite{ott}, of which chaos theory is a part, has
its origins in two famous problems of classical physics: to find a
solution for the three
body problem of celestial mechanics \cite{poincare}, and to find a proof of
Boltzmann's ergodic hypothesis for the equivalence of time and ensemble
averages of dynamical quantities (such as the force per unit area on
the wall of the container) for isolated systems composed of a large
number of particles \cite{lebpen}. In each case a great deal of progress toward an
understanding of both the possibilities and difficulties in finding
solutions to these problems has been made possible, over the past
several decades, by the development
of powerful and sophisticated mathematical and computer methods. These
developments are well summarized in a number of excellent books  and
papers on
dynamical systems theory, some of which are listed as
Refs. \cite{ott,rueck,gasbook,guckho,yorkebo} as well as references
contained therein. This article is a summary of a more extensive set
of notes on this subject developed by the author which presents these
topics in a somewhat more leisurely way, and also contains an
extensive discussion of the Boltzmann transport equation
\cite{jrd}. Another discussion of many of these topics can be
found in 
Ref. \cite{jrdhvb}.

The purpose of this article is to explain some of these developments
and to explore their application to the issues that preoccupied
Boltzmann, namely the foundations of non-equilibrium statistical
mechanics, with a particular emphasis on the foundations of the
kinetic theory of dilute gases of particles that interact with short range
forces \cite{dohvb,resdel,chapcow}. The dynamical properties of such a dilute gas are somewhat
easier to understand than those for other states of matter since in
the main, the dynamics of a particle in a dilute gas consists of
long periods of motion free from interactions with the other
particles, punctuated by occasional and essentially brief collisions with
other particles in the gas. That is, for a gas of particles with short
range forces, the mean free time between collisions is much longer
than the average duration of a collision. Even for such an apparently
simple system complicated and difficult questions arise such as: How
does one reconcile the reversible equations of motion with the
irreversible macroscopic properties of such a system, particularly as
seen in the Second Law of Thermodynamics concerning the irreversible
increase of entropy? Why is the Boltzmann equation, which is
admittedly derived on the basis of stochastic rather than mechanical
assumptions, so successful in predicting the properties of dilute
gases, and in explaining the origin of the hydrodynamic properties of
the gas? and so on. This article will try to show how such questions
might be answered by chaos theory and to indicate the general lines of
the answers, to the extent that the issues are understood at present. Our
discussions will be based at first upon the work of R. Bowen \cite{bowen},
D. Ruelle \cite{rueck,ruelle}, and Ya. Sinai \cite{sinai}, who are the principal architects of the modern
theory of the dynamical foundations of statistical mechanics. Recent
developments due to major contributions of many other authors will
also be described here, with appropriate, though certainly incomplete,
references to the literature \cite{kathas}. 

The plan of this article is as follows: In Section 2 we will review
the basic ideas of ergodicity and mixing for dynamical systems, and
show their importance for statistical mechanics. We discuss the role
of systems with few degrees of freedom, such as the baker's map and
the Arnold cat map, as systems that exhibit ergodic and mixing
properties and serve as paradigms for the kinds of properties we would
like to see in systems with many degrees of freedom. The examples
provided by such simple systems will motivate the important
definitions of Lyapunov exponents, Kolmogorov-Sinai (KS) entropies,
and hyperbolic, Anosov systems which will be important for the further
discussions. In Section 3 we consider some recent applications of ideas
from dynamical systems theory to the theory of transport phenomena,
and illustrate the role of fractal structures that are responsible for
diffusion and other transport processes. We consider there the application of escape-rate
methods to transport theory, as developed by P. Gaspard and
G. Nicolis \cite{gasni}, and the method of thermostatted dynamical systems,
developed by D. Evans, W. Hoover, H. A. Posch, G. Morriss, and
E. G. D. Cohen \cite{thermos}, as well as other authors. In Section 4 we use these
ideas to present our current understanding of the dynamical origins of
the Boltzmann equation, and to show how this equation, in turn, can be
used to calculate quantities such as Lyapunov exponents which measure
the chaotic nature of the relevant dynamics. We conclude in Section 5
with a number of remarks and questions.

 This article is dedicated to the memories of T. H. Berlin, M. S. Green,
M. Kac, J. Kestin, E. A. Mason, P. Resibois, and G. E. Uhlenbeck.

\section{Dynamical Systems Theory and Statistical Mechanics}

We begin by recalling that classical statistical mechanics represents
the mechanical state of a system of $N$ point particles in
$d-$dimensional space by a single point in a
phase space of $2Nd$ dimensions, $\Gamma$-space, whose axes represent
the position and momentum in each spatial direction for each of the $N$
particles \cite{uhlfo}. If we denote a point in $\Gamma$-space by ${\bf X}$, then the
Newtonian motion of this phase point over a time $t$ from ${\bf X}$ to
${\bf X}_t$
will be described by a time displacement transformation $S_t$ with 
\begin {equation}
{\bf X}_t = S_t({\bf X}).
\label{1}
\end{equation}
Although statistical mechanics is usually concerned with systems where $N$
is large and $t$ is continuous, usually called ``flows", here we will
often consider simple dynamical systems in phase spaces of two
dimensions and where the time $t$ takes on
discrete values. Such low dimensional systems with discrete time steps
are called ``maps" \cite{ott,yorkebo}. The reason for considering them is that they display the kind of dynamical properties that we would
like to find in systems of more interest to physicists. In particular
we would like to identify the important features of these simple maps which
are responsible for such properties as ergodic behavior and an
approach to an equilibrium state, and then see if such properties are
characteristic of more general flows for more realistic models of
physical systems.

\subsection{Invariant Measures}
 
Having defined the phase space $\Gamma$ and the time displacement
operator for either continuous or discrete times, we now add a third
basic quantity to our discussion, an invariant measure, $\mu (A)$,
where $A$ is some set in phase space. A measure $\mu$ is invariant
under the time displacement operator $S_t$ if for any set $A$,
$\mu(A)=\mu (S_{-t}A)$, i.e. the measure of the pre-image of $A$,
denoted by $S_{-t}A$ is equal to the measure of $A$ for any
$t>0$. This condition, that an invariant measure
be constant under time evolution, is closely related to the usual
statements in equilibrium statistical mechanics where we append to the
evolution equation for the Liouville distribution function, $\rho({\bf
X},t) = \rho(S_{-t}({\bf X},0)$, the requirement that $\rho$ be
independent of time. An example of the  
point of the requirement that $t > 0$, will be seen shortly when we
consider invariant measures for one dimensional maps, which are not invertible. Well known
examples of invariant measures include the volume of a set in
$\Gamma$-space which defines the Liouville measure, $\mu_{L}(A)$,
\begin{equation}
\mu_{L}(A) = \int_{A}d \Gamma,
\label{2}   
\end{equation}   
where the integration is over the range of position and momentum
coordinates that define the set $A$ in $\Gamma$-space; the microcanonical measure on
the constant energy surface, $\mu_{m.c.}(A)$, defined by
\begin{equation}
\mu_{m.c.}(A) = \int_{A}\frac{d\Omega}{|\nabla H|},
\label{3}
\end{equation}
where $d \Omega$ is a surface element on the microcanonical energy
surface, and $\nabla H$ is the gradient of the Hamiltonian $H$ of the
system with respect to all of the coordinates and momenta in
$\Gamma$-space \cite{uhlfo}. 

There are a number of useful systems of low dimensionality with
simple invariant measures, as well. As they will be used in our
further discussions, we describe the systems and their appropriate
invariant measures now. We should mention that very often dynamical systems have
infinitely many invariant measures. However of these, usually only one, called the natural or SRB measure, can be used to compute the type of ensemble
averages used in statistical mechanics. The other measures are often
defined on very special points or sets in phase space. For example,
one can define invariant
delta function measures on periodic points of the system, but these
measures can not generally be used to characterize the statistical
properties of the system. In all of the disussions to follow we will
assume that the total measure of the appropriate phase space is
finite, which implies, of course, that the measure of any subset of the space is finite. 

The simplest map with an invariant measure is a piecewise linear, 1-d non-invertible map of the
unit interval, $[0,1]$ {\it onto} itself, an example of which is the tent map
illustrated in Fig. 1. This map is given by \cite{ott,yorkebo}
\begin{equation}
\begin{array}{ccccc}
x_{n+1} & = & 2x_n \,\, & {\rm for} & \,\, 0 \leq x_n < 1/2, \\
& = & 2(1 -x_n)  \,\, & {\rm for} & \,\, 1/2 \leq x_n \leq 1
\end{array}
\label{4}
\end{equation}
Here the invariant measure of a set such as $A$, $\mu (A)$
illustrated in the figure is its Lebesgue measure, or simple area, $\mu_{l}(A)$. For this map the
pre-image of $A$ is the union of two sets each of Lebesgue measure
$\mu_{l}(A)/2$. Note that the image of $A$ typically has Lebesgue measure
$2\mu_{l}(A)$, so our definition of the invariant measure, using the
condition $t>0$ above allows us to define such measures even for
non-invertible maps, such as the tent map. an invariant measure should be defined by
considering the pre-image of $A$, as we did above. 

For our purposes
the most useful maps will be two dimensional ($2-d$) invertible maps,
particularly, the baker's map illustrated in Fig. 2, and the toral
automorphism, often called the Arnold cat map, illustrated in
Fig. 3. \cite{arnav}. The baker's map is defined on the unit square $0\leq x,y \leq
1$ by the dynamical equation
\begin{eqnarray}
\left(\begin{array}{c}
x_{n+1} \\
y_{n+1} \end{array}\right) & = & {\bf B}\left(\begin{array}{c}
x_n \\
y_n 
\end{array}\right) \nonumber \\
&= & \left( \begin{array}{c}
2x_n \\
y_{n}/2 \end{array}\right) \,\, {\rm for} \,\,0\leq x <1/2, {\rm and}
\nonumber \\
& = & \left(\begin{array}{c}
2x_{n}-1 \\   
(y_{n}+1)/2 \end{array}\right)\,\, {\rm for} \,\,1/2\leq x \leq 1.
\label{5}
\end{eqnarray}
This map has an invariant measure which is just the Lebesgue measure
of two-dimensional regions on the unit square. This follows
immediately from the observation that horizontal lengths become twice
as long, but vertical lengths become one half as long. One can also
define the inverse map ${\bf B}^{_1}$ in a simple way. Under the
inverse map, horizontal lengths contract by a factor of $2$, and
vertical lengths become twice as long. 

Toral
automorphisms, ${\bf T}$, are defined in general by the transformation
\begin{equation}
\left( \begin{array}{c}
x_{n+1}\\
y_{n+1} \end{array} \right)
= {\bf T}\cdot \left( \begin{array}{c} x_n \\ y_n \end{array} \right )
= \left( \begin{array}{cc}
a & b \\
c& d \end{array}\right) \cdot \left(\begin{array}{c}
x_n \\
y_n \end{array}\right) \,\,\,\,{\rm mod}\,\,1,
\label{6}
\end{equation}
where $a,b,c,d$ are positive integers, and the determinant of the
$2\times 2$ matrix, $ac-bd =1$. Thus a unit torus is mapped onto a
unit torus, and the condition on the determinant guarantees that all
areas on the unit torus are preserved by the mapping. For reasons to be
clear shortly, we will restrict
our attention here to so-called hyperbolic maps where the eigenvalues
of ${\bf T}$ do not lie on the unit circle. The Arnold cat
map is a particular case of this where the matrix ${\bf T}$ is given
as $\left ( \begin{array}{cc}
2 & 1 \\
1 & 1 \end{array}
\right).$ This is the case illustrated in Fig. 3.

\subsection{Ergodic and Mixing Systems}

For classical systems, the foundations of equilibrium statistical
mechanics are based upon Boltzmann's {\it ergodic hypothesis}: The
long time average of the dynamical quantities accessible to
measurement, such as the force per unit area exerted by a fluid on the
walls of a container, for an individual, isolated system, is equal to
the ensemble average of the same dynamical quantity taken with respect
to a microcanonical ensemble \cite{uhlfo}. One of the tasks of dynamical systems theory is
to identify the kinds of dynamical systems and dynamical quantities
for which the ergodic hypothesis can be verified. 

The first major
result in this direction was Birkhoff's ergodic theorem \cite{arnav,birkh}: Consider some
dynamical function, $F({\bf X})$, defined on the constant energy
surface, ${\cal E}$, that satisfies the condition
\begin{equation}
\int d\mu F({\bf X}) < \infty,
\label{7}
\end{equation}
where the integration is with respect to the microcanonical measure
defined in Eq. (\ref{3}), and is carried out over the whole
surface. Suppose further that the total measure of the surface is
finite. Then: (i) the time average of $F$, denoted by $\bar{F}({\bf
X}) = \lim_{T \rightarrow \infty}\frac{1}{T}\int_0^{T} dt F({\bf
X}(t)) $ exists almost everywhere on the constant energy surface, that
is, for almost every starting point ${\bf X}$; (ii) The time average
$\bar{F}({\bf X})$ may depend on the particular trajectory but not upon
the initial point of the trajectory; (iii) The ensemble average of
$\tilde{F}$ is equal to ensemble average of $F$ itself. That is,
$\int d \mu F({\bf X}) = \int d \mu \bar{F}({\bf X})$; and (iv) One can
replace the entire constant energy surface as the region of
integration by any invariant subset of non-zero measure. Here an
invariant set is one where all points (with the exception of a subset
of measure zero) initially in the set remain in the set during the
course of their time evolution, and the measure of the set does not
change with time. Birkhoff's theorem does not show that systems of
physical interest are ergodic in the sense of Boltzmann because the
time average of an integrable dynamical quantity may depend on
trajectories and thus not necessarily be equal to an ensemble average,
which is, of course, just a number. 
We therefore define an ergodic
system to be one where the time average $\bar{F}$ of any dynamical quantity $F$
is a constant on the surface. For such a system the time average of
$F$ is indeed its ensemble average, 
 which we denote by $\tilde{F}$,
since
\begin{equation}
\int d \mu F({\bf X}) = \bar{F} \int d \mu,
\label{8}
\end{equation}
from which it follows that 
\begin{equation}
\bar{F} = \frac{\int d \mu F({\bf X})}{\int d \mu} = \tilde{F}.
\label{9}
\end {equation}

It should be remarked that although it is true that for ergodic
systems the ensemble average of any integrable function is equal to
its time average, the requirement of ergodicity may be demanding too
much of the dynamical systems to which we would like to apply
equilibrium statistical mechanics. For example, for applications of
statistical mechanics we might restrict ourselves to a smaller class
of functions for which we would need to prove the equivalence of time
and ensemble averages, thus placing fewer demands on the dynamics.

Over the seventy or so years since Birkhoff's theorem was proved, a
number of systems have been shown to be ergodic \cite{sinai,kathas,szasz}. These include simple
transformations of the unit interval $[0,1]$ onto itself by $x' =
x + \alpha$ mod $1$, (think of taking steps of length $\alpha$ on a
circle of unit circumference) where $\alpha$ is irrational, the baker's map,
toral automorphisms of the type described above, geodesics on surfaces
of constant negative curvature, and of most interest for physics,
billiard ball systems in two and three dimensions. These systems
include a particle moving among a system of fixed, non-overlapping
hard disk or hard sphere scatterers (the Lorentz gas) \cite{sinai}, as well as
certain systems of moving hard disks or hard spheres, in a box with periodic
boundary conditions \cite{szsim}. The proofs of ergodicity for billiard systems
are not at all easy(!), but billiards may be the simplest physically
realistic systems to analyze because of the simplicity of the
collisions of the particles \cite{livwojt}. We will return to this point further on,
but first we wish to develop further characterizations of the
dynamical systems for which the methods of statistical mechanics
apply.

Gibbs took Boltzmann's ideas a step further by introducing the idea of
a {\it mixing} system \cite{kathas,uhlfo,arnav}. He proposed to look at an initial ensemble of
mechanically identical systems all with the same total energy, but with
different initial conditions. Suppose the phase points for this
ensemble are confined, initially to a set of positive measure, $A$, on the constant energy
surface, ${\cal E}$. Then in the course of time this set will evolve
to a new set, $A_t$, with the same measure as $A$. A system is called
mixing in the sense of Gibbs, if for any fixed set $B$ of positive
measure on ${\cal E}$
\begin{equation}
\lim_{t \rightarrow \infty}\frac{\mu (B \cap A_{t})}{\mu (B)} =
\frac{\mu (A)}{\mu ({\cal E})}.
\label{9}
\end{equation}
That is, a system is mixing if the time
evolution of any set of non zero measure leads to the set being
distributed uniformly with respect to the invariant measure $\mu$ over the constant energy surface. 

It is easy to show that the condition that a dynamical system be
mixing is stronger than requiring that it be ergodic \cite{kathas,arnav}. That is, all
mixing systems are ergodic but not {\it vice versa}. For example
the ergodic transformation of the unit interval $[0,1]$ onto itself, by
$x' = x + \alpha$ mod $1$, for irrational $\alpha$ is not mixing. To see
this just let $A, B$ each be small connected sets of the unit interval, and
consider the set $C_n=A_{n}\cap B$, the intersection of $B$ with the
image of $A$ after $n$ applications of the
transformation. The limit of $\mu (C_n)$ as $n$ gets large is not well
defined and the set $A$ does not get uniformly mixed over the
unit interval, but instead moves as a rigid set without mixing.

The mixing property of a dynamical system is even more interesting for
our purposes than that of ergodicity, not only because mixing implies
ergodicity, but also because if a system is mixing, then
non-equilibrium ensemble averages approach their equilibrium values as
time gets large \cite{jrd,arnav}. That is, if one starts with some non-equilibrium
ensemble distribution on ${\cal E}$, not the microcanonical one, and
if this distribution is a measureable function on ${\cal E}$, then the
time evolution of this distribution function is governed by
Liouville's equation. The ensemble average of any measurable dynamical
quantity, $F$, at some time $t$ is then determined by the integral of
$F$ with respect to the distribution function at time $t$. For a
mixing system, this ensemble average approaches its value in the
microcanonical ensemble as $t$ approaches infinity. Physically this
means that in a weak sense, that is, under integration with some well
behaved function , non-equilibrium distribution
functions approach the microcanonical equilibrium distribution in the
long time limit. This is the essential information needed for
the validity of non-equilibrium statistical mechanics.

Systems that have been rigorously proved to be mixing include the
baker's map, the hyperbolic toral automorphisms, and the Lorentz gases described
above \cite{kathas,gallor,sinai2}. There is little doubt the moving hard sphere, or hard disk
systems are mixing, but as yet there is no rigorous proof of this
supposition.
 
\subsection{Dynamics, Symbolic Dynamics, and Lyapunov Exponents}

While we will not prove here that the baker's map or the toral
automorphisms are mixing systems, it is very instructive to sketch
the lines of the proof for the baker's map. This will provide a simple
example of a {\it Markov partition}, which is a useful construction in
many other, and more general, contexts \cite{peterson,billings}. We begin by noting that any
number in the interval $[0,1]$ can be expressed in a binary series,
the dyadic expansion, as
\begin{equation}
x = \frac{a_0}{2}+\frac{a_1}{2^2}+\frac{a_2}{2^3}+\cdots,
\label{11}
\end{equation}
where each of the $a_i$ can take on the values $0,1$, depending on
$x$, of course. We can represent $x$ as an infinite sequence of
$0$'and $1$'s as
\begin{equation}
x=(a_0,a_1,a_2,\cdots).
\label{12}
\end{equation}
This representation is unique except for those fractions whose
representations consist of a finite number of $a$'s followed by an
infinite number of $0$'s or of $1$'s. Such fractions have two
equivalent representations, one with an infinite sequence of zeroes
and one with an infinite sequence of $1$'s. As these particular
fractions form a countable set, we can always require that we use one of
these representations for them consistently. In a similar manner we may
represent $y$ in $[0,1]$ as
\begin{eqnarray}
y & = & \frac{b_{-1}}{2}+\frac{b_{-2}}{2^2}+\cdots\,\, {\rm or} \nonumber
\\
y & = & (b_{-1},b_{-2}, \cdots).
\label{13}
\end{eqnarray}
Combining these two representations, we see that any number in the
unit square $0\leq x,y \leq 1$ may be expressed by
\begin{equation}
(x,y) = (\cdots b_{-2}b_{-1}.a_{0}a_{1}a_{2}\cdots),
\label{14}
\end{equation}
where we have separated the $x$-sequence from the $y$-sequence by a
``dot", and strung the $y$-sequence to the left, with the $x$-sequence to
the right. We can clearly approximate any point in the unit square to
an accuracy $2^{-N}$ by a finite sequence
\begin{equation}
(x,y) \simeq (b_{-N}b_{-N+1}\cdots b_{-1}.a_{0}a_{-1}\cdots a_{N-1}).
\label{15}
\end{equation}
Such an approximation is equivalent to locating the point somewhere in
one of the small squares obtained by dividing the unit square into
small regions by drawing vertical lines at intervals of $2^{-N}$, and
similarly with horizontal lines. This partition of the unit square
into much smaller regions, is an example of a Markov partition,
because of the particular way in which the lines are drawn, as we
explain below. The approximation to the
point $(x,y)$ consists in specifying the particular small region in which the
point is located. This procedure for approximating a particular point
in phase space, here the unit square, 
is a mathematical echo of the idea of coarse graining
often used in statistical mechanics.

The central feature that we now want to emphasize is that given the
representation of the point $(x,y)$ on the unit square by
Eq. (\ref{14}), its image, $(x',y')$ under the baker's map becomes
\begin{equation}
(x',y') = (\cdots b_{-2}b_{-1}a_{0}.a_1a_2 \cdots),
\label{16}
\end{equation}
where all the members of the sequence have shifted one unit to the
left, and, in particular, the number $a_0$ has moved past the ``dot"
into the $y$-sequence. Thus we have mapped a deterministic dynamical
system onto something that looks like a ``coin toss"
experiment. A point on the unit square can be mapped onto a bi-infinite sequence of
$0$'s and $1$'s, that is, onto a particular realization of a
bi-infinite coin toss sequence, and the ``dot" indicates the location
of one particular moment in the sequence of tosses. The next toss will
simply be the next step in the process of generating the entire
sequence, and corresponds, in the baker's map, to one iteration of the
map! A coin toss experiment is, of course, a stochastic process, with
a given probability $p$ for obtaining a ``heads" and $q=1-p$ for a
``tails". Our ability to map certain deterministic dynamical
systems onto random, stochastic processes is the main point of this
article! Of course the central question is to identify those
systems for which such a mapping is possible.

Now consider what happens to all of the points in the small square of
order $2^{-N}$ on a side, which we will take to be the set $A$ in the
Gibbs picture. We know that for all such points the values of
$b_{-N},b_{-N+1}, \cdots , b_{-1}, a_0, a_1, \cdots, a_{N-1}$ will be
the same. However, each of the additional, unspecified $a_{N+k},
b_{-N-k'}$, with $ k>-1, k'>0$, can take on the values $0$ or $1$. All of the
points in the small square correspond to all possible values of these
unspecified $a$'s and $b$'s. After $N$ iterations of the baker's map, all of the
specified values for the $a$'s will have shifted onto the $y$
coordinate, and the set of points in the original small square will be
uniformly distributed along the $x$ interval $[0,1]$. After $2N$
iterates of the map, we will have lost track of the $y$ coordinates as
well, at least to the accuracy we have specified, and the points
originally confined to our small square will be uniformly distributed
over the entire unit square. Thus the system is indeed mixing. Note
also that a similar circumstance would apply if the map were run
backward, and the sequence shifted to the right past the ``dot" at
each step. Thus we can see an approach to a uniform distribution for both the
``forward" and the ``time reversed" motions. Consequently we can say
that it is certainly possible for the ensemble distribution for a reversible dynamical system to show an approach to a uniform,
equilibrium state for both forward and backward motions, provided we
look at the distribution on a slightly coarse grained scale. 

A very
similar analysis applies to the hyperbolic toral autormophisms, but the
partitioning of the unit square (or torus) should be done with lines
whose directions and spacings are dictated by certain features of the
particular automorphism \cite{kathas,peterson}. These features will be determined by the
directions of the stable and unstable manifolds, and by the Lyapunov
exponents of the map, all of which we will discuss shortly.
The representation of the trajectory of a point
on the unit square under the baker's map in terms of sequences of
$0$'s and $1$'s, is an example of {\it symbolic dynamics}, whereby,
trajectories are coded into sequences whose elements are chosen from a
finite number of symbols (here $0$ or $1$) \cite{ott,yorkebo,kathas}. The representation of
dynamics using such sequences can be very useful in analyzing and
calculating the properties of dynamical processes in simple enough
systems. For example, readers might convince themselves that the
Cantor ``middle third" set can be coded by similar sequences of two
symbols. (Hint: Consider the map $x' = 3x$ mod $1$, and represent all numbers on the unit interval by a series
in inverse powers of $3$, with coefficients $0,1,2$. The
``middle third" Cantor set is isomorphic to all sequences where a $1$ never
appears, and the above transformation just maps this set onto itself.)

Now a remarkable thing has just happened in our analysis of the baker
map. We started with a set of points that were all within a distance
$\epsilon = 2^{-N}$ of each other, and they spread out over the
unit square. Consequently, the rates and directions of separation, or
conversely of approach, of the images of two nearby points are certainly
quantities of some interest for understanding the mixing process. For
the baker's map, it is clear that the images of infinitesimally close
points with the same $y$ coordinate will separate in the
$x$-direction, so that their separation, $d_n$, after $n$ steps will
be $2^{n}d_0$, where their initial infinitesimal separation is
$d_0$. However, the images of two infinitesimally close points with the same $x$
coordinate will approach each other as $d_n=2^{-n}d_0$. The images of
two infinitesimally close points that do not lie in a strictly
vertical line will also separate exponentially, since the exponential
separation in the $x$ - direction will dominate the separation of the
two points. Consequently, the images of any two infinitesimally close
points will separate exponentially, except for a set of measure zero,
namely those points with the same $y$ coordinate. If we consider the
time reversed motion, then points will separate in the $y$ direction,
approach in the $x$- direction, but the images of almost all nearby
points will separate exponentially rapidly. The exponents that govern
the rates of exponential separation or approach are called {\it
Lyapunov exponents} \cite{ott,rueck}. We see that the baker's map has two Lyapunov
exponents, $\ln2, -\ln2$. The fact that the sum of these exponents is
zero is not accidental, it is a consequence of the invariance of the Lebesgue
measure on the unit square. Moreover, while the separation of the
nearby points is exponential for a large number of iterations of the
map and the rate of separation is given by the Lyapunov exponents on
this time scale. However this exponential rate of separation cannot continue
forever since the measure of the phase space is finite.. Eventually the separation will become large enough that the
folding mechanisms of the baker map will take over and the two points
will find themselves in very different regions of the unit square. In
general we will find that exponential separation of trajectories in a
bounded phase space cannot continue indefinitely, but that there is
always a folding mechanism that keeps trajectories within the bounded region.   

Consider now the case of the hyperbolic toral automorphisms, Eq. (\ref{6}), and let us use the
Arnold cat map as an example. An elementary calculation shows that the
$2\times2$ matrix given below Eq. (\ref{6}) has the eigenvalues
$\Lambda_{\pm}$ given by
\begin{equation}
\Lambda_{\pm} = \frac{3\pm 5^{1/2}}{2}.
\label{17}
\end{equation}
Notice that $\Lambda_{+}>1, \Lambda_{-}<1$, and the invariance of the
Lebesgue measure on the unit torus requires that
$\Lambda_{+}\Lambda_{-}=1$. Here again we see that there is an
expanding direction which is determined by the direction of the
eigenvector corresponding to $\Lambda_{+}$, and a contracting
direction determined by the direction of the eigenvector corresponding
to $\Lambda_{-}$. The associated Lyapunov exponents which we denote by
$\lambda_{\pm}$ are given by $\lambda_{\pm}=\ln \Lambda_{\pm}$. Here
again we note that the invariance of the measure on the unit torus
requires that $\lambda_{+} + \lambda_{-}=0$. Consider a typical point
on the unit square and draw a set of lines through it such that one is
in the direction of the expanding eigenvector and the other is in the
direction of the contracting eigenvector. We will refer to these two
lines as the {\it expanding or unstable manifold} and the {\it contracting
or stable manifold} respectively. The images of two infinitesimally close points on
the unstable manifold will separate exponentially, while the images of
two close
points on the stable manifold will approach exponentially
\cite{ott,kathas}. 

Now we can return to the idea of a Markov partition
mentioned at the beginning of this section. A Markov partition is a
partition of the phase space into a finite collection of small ``parallelograms" with
disjoint interiors and whose sides lie
on stable and unstable manifolds, perhaps suitably extended, of points
in the space. For the case of the baker's map, the parallelograms have
vertical and horizontal sides, while for the Arnold cat map, the
parallelograms have sides that lie along the eigen directions of the
matrix below Eq. (\ref{6}). The existence of a Markov partition for a
dynamical system allows one to use  symbolic dynamics to code the
trajectories, and to use of the theory of Markov processes to
analyze the time dependence of distribution functions defined on the
phase space. For example, the solution of the Perron-Frobenius
equation to be discussed in Section 4 is greatly facilitated if
one can construct a suitable Markov partition for the system. The
calculations described in Section 3.4 makes extensive use of the
method of Markov partitions. Further details and a discussion of the
use of Markov partitions to compute other quantities of dynamical
interest such as Kolmogorov-Sinai entropies, to be discussed below,
can be found in Refs. \cite{sinai,kathas}. 

  In order to show that the
Lyapunov exponents are non-zero for both the baker map and the
hyperbolic toral
automorphisms we have had to calculate them, thus, given a dynamical
system, it is not always obvious that it will have the positive
Lyapunov exponents that are required for the exponential separation of
trajectories.

We can generalize the concept of Lyapunov exponents, stable, and
unstable manifolds to higher
dimensional systems and to flows, in addition to maps \cite{rueck}. Usually the
only fundamental difference between maps and flows is that flows
typically have one zero Lyapunov exponent in the direction of the
trajectory in phase space \cite{gasbook,rueck}. Although the proof can sometimes be
involved mathematically, the zero Lyapunov exponent in the direction of
the trajectory can be understood by imagining two infinitesimally
close points that lie on the same trajectory in phase space. If the
dynamics of the system consists of free particle motion punctuated by
collisions that are instantaneous or almost so in comparison with
other time scales of the system (the mean free time between
collisions, say), then the two points will remain close over the course
of their motion and their separation will certainly not be
exponential. Furthermore, if the dynamics is consistent with an
invariant measure, such as Hamiltonian dynamics, then the sum of {\it
all} the non-zero Lyapunov exponents must be zero as a consequence of
the invariance of the measure. This suggests that the Lyapunov
exponents should be defined in such a way as to be consistent with the
invariance of the measure, which is the case in the two examples we
have considered. It is worth pointing out that Lyapunov exponents for
periodic trajectories, which have delta function invariant measures,
may very well be different from those defined with respect an
invariant measure defined on the entire phase space. One additional property of the non-zero Lyapunov
exponents for a Hamiltonian system that is important is the {\it
symplectic conjugate pairing rule}. That is, for a system with
symplectic dynamics, the non-zero Lyapunov exponents, should there be
any, must come in ``plus-minus" pairs, with each pair of exponents
summing separately to zero, independent of the values of the other
pairs. This is a consequence of the symplectic form of Hamilton's
equations of mechanics \cite{arnold}, and is not true for reversible systems that
are not symplectic. We shall encounter such systems in the further sections. 

\subsection{The Kolmogorov-Sinai Entropy}

Let us now look at some properties of the Arnold cat map in more
detail. In Fig. 4, we show an initial set $A$ which is located in the
lower right hand corner of the unit square, and in Figs. 5, 6, 7, we
show the evolution of this set after 2, 3, and 10 iterations of the
map respectively (a similar set of figures can be found in \cite{lasmak}). As the number of iterations increases the set becomes longer and thinner such that at the third iteration the set has begun to
fold back across the unit square, and after 10 iterations the set is
so stretched and folded that it appears to cover the unit square
uniformly. If we were able to increase the resolution of this
illustration beyond the $10^5$ points used to generate it, we would see
that this apparently uniform distribution is made up of very many, on
the order of $10^6$, thin
lines  parallel to the expanding direction of the cat map, and very
close together.  Since the initial square is getting stretched along
the unstable manifold, at every iteration we learn more about the
initial 
location of points within the small initial set $A$. That is, suppose
we can distinguish two points on the unit square only if they are
separated by a distance $ \delta $, the resolution parameter, and
suppose further that the initial set has a characteristic dimension on
the order of $\delta$ \cite{jrdhvb}. We cannot resolve two points in the initial set
then, but after a time t, the initial set will have stretched along
the unstable direction to a length on the order of $\delta \exp
(\lambda_{+}t)$, and we can easily resolve the images of points in the
initial set. Thus as we look at the successive images of the initial
set we are able to obtain more and more information about the location
of points in the initial region, in fact, the information is growing
at an exponential rate. The exponential rate at which information is
obtained is measured by the {\it Kolmogorov-Sinai (KS)
entropy} \cite{ott,rueck,kathas} , $h_{KS}$, and for our
simple system, the cat map, $h_{KS}=\lambda_{+}$. In general, one
finds that for a dynamical system of several dimensions, with positive
Lyapunov exponents, and where all of the
points are confined to a bounded region of phase space, then
\begin{equation}
h_{KS} = \sum_{i}\lambda_{+,i},
\label{18}
\end{equation}
where the summation is over all of the positive Lyapunov exponents of
the system. Eq. (18) is referred to as Pesin's theorem
\cite{rueck,pesin}, and we mention
that there is a highly developed theory for the KS entropy \cite{kathas,peterson}, which
we cannot expand upon here. We mention again that the Lyapunov
exponents and the KS entropy should be defined with respect to an
invariant measure. Thus, there may be many sets of exponents for a
single system, e.g., Lypaunov exponents defined on
periodic orbits may differ from those defined with respect to the
natural invariant, or SRB, measure, which will be discussed in Sec. 2.5. Pesin's theorem applies if $h_{KS}$ and the
Lyapunov exponents are defined with the respect to the natural
invariant measure. In the next section we will consider a
situation where points can escape from a bounded phase space region, and where
the KS entropy is less than the sum of the positive Lyapunov
exponents by an amount that is equal to the escape rate of points from
the bounded region. One might say that in this latter case the rate of
stretching is
slightly greater than the rate of folding. 
 
\subsection{Hyperbolic and Anosov Systems}
 
We have argued that both the baker's map and the hyperbolic toral automorphisms
are mixing systems, and as a consequence, as simple models of
dynamical systems, they show an approach to equilibrium. In fact we
can map these simple reversible dynamical systems onto stochastic
processes. The mechanisms responsible for this desirable behavior are:
(1)
the exponential separation of trajectories in phase space,
characterized by positive (with corresponding negative) Lyapunov
exponents with corresponding expanding and contracting directions; and
(2) the folding of phase space regions during their time evolution due
to the boundedness of the phase space, leading to a uniform
distribution of points over the phase space, at least on a coarse
grained scale. It is now time to formalize these properties so that we
can see if the above mechanisms apply for more realistic systems. 

To do this we consider a general dynamical system with several degrees
of freedom, with an associated phase space $\Gamma$ and an invariant
measure $\mu$ on the phase space. Let us write the equations of motion
for the phase space variables, denoted by ${\bf X}$ as
\begin{equation}
\dot{\bf X} = {\bf G}({\bf X}),
\label{19}
\end{equation}
where, typically, but not always, ${\bf G}$ is the appropriate derivative of the
Hamiltonian function with respect to the phase space variables.
In order to consider the possible exponential separation
of infinitesimally nearby trajectories, we consider the time evolution
of a small displacement $\delta{\bf X}$ in phase space between two
arbitrarily close points \cite{gasbook,gasdo}. This displacement satisfies a linearized
equation in the tangent space to $\Gamma$ at the point ${\bf X}$,
\begin{equation}
\dot{\delta{\bf X}}=\frac{\partial{\bf G}({\bf X})}{\partial{\bf
X}}\cdot \delta{\bf X}.
\label{20}
\end{equation}
The solution to this equation in the tangent space has the form
\begin{equation}
\delta{\bf X}_{t}={\bf M}(t, {\bf X})\cdot {\bf X},
\label{21}
\end{equation}
where ${\bf M}$ satisfies the equation
\begin{equation}
\dot{{\bf M}}(t, {\bf X})= \frac{\partial{\bf G}({\bf X}_t)}{\partial{\bf
X}_t}\cdot {\bf M}(t,{\bf X}).
\label{22}
\end{equation}
The matrix ${\bf M}$ plays a special role in the discussion now, as it
determines the dynamical evolution of the displacements in the
tangent space.

We define an {\it Anosov} dynamical system as a map or flow with the
following properties \cite{guckho,kathas,lanford} (we consider flows here, but our definition can
be easily modified for maps):

1) for almost every point, ${\bf X}$, in the phase space $\Gamma$, there is
a decomposition of the tangent space at ${\bf X}, {\cal T}\Gamma ({\bf X})$, into
three subspaces, an unstable subspace ${\cal E}_{X}^{u}$, a stable
subspace ${\cal E}_{X}^{s}$, and a center subspace ${\cal E}_{X}^n$
such that
\begin{equation}
{\cal T}\Gamma (X) = {\cal E}_{X}^{u} \oplus {\cal E}_{X}^s \oplus {\cal E}_{X}^0.
\label{23}
\end{equation}

2)There are constants $C_s, C_u$, and $\Lambda$ with $C_s, C_u >0$, and $ 0< \Lambda <1$, such that:

a) If $\delta {\bf X}$ is in ${\cal E}_{X}^s$, then
\begin{equation}
||{\bf M}(t,{\bf X}) \cdot \delta {\bf X}|| \leq C_{s}\Lambda^{t}||\delta
{\bf X}||,
\label{24}
\end{equation}

b) If $ \delta {\bf X}$ is in ${\cal E}_{X}^u$, then (for $t>0$)
\begin{equation}
||{\bf M}(-t,{\bf X}) \cdot \delta {\bf X}|| \leq C_{u} \Lambda^{t}||\delta{\bf X}||,
\label{25}
\end{equation}

c)The subspaces ${\cal E}_{X}^u, {\cal E}_{X}^u, {\cal E}_{X}^0$ vary
continuously with ${\bf X}$. This means that any vector $\delta{\bf
X}$ can be written as 
\begin{equation}
\delta {\bf X}=\delta {\bf X}_{X}^u +\delta {\bf X}_{X}^s +\delta {\bf
X}_{X}^0,  
\label{26}
\end{equation}
where each term is in the indicated subspace. Then each of the $\delta
{\bf
X}_{X}^{j}$ vary continuously with ${\bf X}$.

d)The various subspaces intersect transversely, without any tangencies.

The center subspace contains the directions of motion with zero
Lyapunov exponent, such as the direction tangent to the trajectory of
the system in phase space, and any other directions associated with
the macroscopically fixed constants of motion, such as the total
momentum, etc. Note that in condition b) above we
considered the time reversed motion on the unstable manifold, and the
condition is that small deviations from the orbit converge
exponentially to zero for the time reversed motion, while for the
stable manifold, the deviations approach zero exponentially in the
forward direction. The transversality condition d) assures that there
are no tangencies among the subspaces which would greatly complicate
the description of the dynamics.

As we will see in the next section, there are a number of
circumstances of physical interest where we will need to consider
invariant subregions of the full phase space. These regions are
usually fractal attractors or repellers, to be defined and described
in the next section, which have Lebesgue measure zero in the full phase
space. It may happen that the dynamics, when restricted
to such an invariant subregion still satisfies conditions a)-d) above. In
such a case we say that the dynamics is {\it hyperbolic} on the
subregion. An Anosov system, then, is one which is hyperbolic on the
full phase space.

Anosov systems of finite total measure have the properties that we need for equilibrium and non-equilibrium statistical
mechanics, they are ergodic and mixing, with positive KS
entropy. However with the exception of the simple models we have
discussed already, we do not know if some of the most physically
realistic systems are Anosov. Certainly hard disk or hard sphere
systems are not Anosov, since the collision dynamics produces
discontinuities in which one member of a pencil of nearby
trajectories just includes a grazing collision involving two
particles, but many of the other trajectories miss that collision
altogether. However, Sinai and co-workers have shown that such systems
still have many of the nice properties of Anosov systems\cite{sinai,livwojt}. In view of
the difficulty in establishing whether a physical system is Anosov,
Gallavotti and Cohen \cite{galco} have proposed a {\it chaotic hypothesis}, namely,
that until one proves that a system of interest is not Anosov, one
should assume that it is, and then calculate its dynamical properties
based upon that supposition. Their reasoning is simply that had we
waited for someone to prove that a given physical system is ergodic
before applying the methods of equilibrium statistical mechanics to
it, there would have been precious little progress in understanding
the properties of equilibrium systems over the last hundred
years. Thus it would seem practical for physicists, at least, to
compute the dynamical properties of systems by assuming that the
systems are Anosov. We will adopt this attitude here, for the most
part, but we will also treat the hard disk Lorentz gas in the later
sections, and make explicit use of the discontinuous dynamics. We'll
see that the failure of this system to be Anosov is not a serious
handicap.
 
We conclude this section with the statement of an important theorem
due to Sinai, Ruelle, and Bowen, \cite{guckho} that is a generalization of the
ergodic theorem to general hyperbolic systems, whether they be an
Anosov system, or a system which is restricted to a repeller or an
attractor. Suppose then that we have: (1) a hyperbolic system described by a
flow ${\bf X}_t = S_{t}({\bf X})$ for almost all ${\bf X}$ in some
invariant region ${\cal R}$, and (2) $g({\bf X})$ is a continuous
function of ${\bf X}$. Then there exists a unique measure $\mu$ such
that for almost all (with respect to Lebesgue measure) ${\bf X}$ in
${\cal R}$
\begin{equation}
\lim_{T \rightarrow \infty}\frac{1}{T}\int_0^T dt g(S_t({\bf X}))
=\frac{\int g({\bf Y})d \mu}{\int d \mu},
\label{27}
\end{equation}
where the integrals on the right hand side of Eq. (\ref{27}) are over
the invariant region ${\cal R}$. The measure that is proved to exist by this SRB theorem is called a
SRB measure. In the case that ${\cal R}$ is the full phase space
$\Gamma$, then the SRB measure is the microcanonical measure, and the
SRB theorem is equivalent to the statement that Anosov systems are
ergodic. However the theorem is much more general than that, and as we
have said, it applies to hyperbolic attractors and repellers, as
well. 

\section{Applications of Dynamical Systems Theory to Transport:
Repellers and Attractors}

\subsection{The Escape-Rate Method}

In our discussion of the KS entropy, we argued, in effect, that if we
have a closed hyperbolic or Anosov system, the sum of the positive
Lyapunov exponents is equal to the KS entropy (Pesin's theorem,
Eq. (\ref{18})). Suppose now that we have a situation frequently
encountered in transport phenomena, where particles can escape from the
system. Consider, for example, the diffusion of particles in a fluid,
that has absorbing boundaries. In that case, particles that reach the
boundary will be lost from the system. Here we argue that it is possible to apply
dynamical systems theory to such a case, and that this application
shows a deep connection between transport coefficients and the
properties of chaotic dynamical systems. This connection was first made by Gaspard
and Nicolis \cite{gasni}.  This is an appropriate place to say that by a chaotic
dynamical system we mean one that has a positive KS entropy, i.e.,
exponential separation of trajectories, and a bounded phase space so
that folding takes place.

First we consider the macroscopic description of such a diffusion
phenomena. Suppose we have some macroscopic region of characteristic
size $L$, much greater than any of the microscopic lengths that
characterize the system. Suppose that the system consists of one
moving particle and a collection of fixed scatterers, arranged in such
a way that the moving particle is never trapped inside the system. Now
suppose that the boundary of the system is absorbing, so that any
particle that reaches the boundary will be absorbed, and removed from
the system. The macroscopic description of this process is through the
Fokker-Planck equation for $P(\vec{r},t)$, the probability of finding
the particle at point $\vec{r}$ at time $t$, which is
\begin{equation}
\frac{\partial P(\vec{r},t)}{\partial t} = D \nabla^{2} P(\vec{r},t),
\label{28}
\end{equation}
supplemented by the absorbing boundary condition, $P=0$ on the
boundary. (In this case the Fokker-Planck equation coincides with
thediffusion equation.) Here the quantity of interest is the diffusion coefficient $D$. If we solve this equation and ask for the probability that
the particle will still be in the system at time $t$, $P(t) =\int
d\vec{r}P(\vec{r},t)$ we find that $P(t)=\exp[-\gamma t]$, where the
{\it escape-rate} $\gamma$ is
\begin{equation}
\gamma = D\frac{a}{L^2},
\label{29}
\end{equation}
where $a$ is a numerical constant that depends on the geometry of the
region of scatterers.

A remarkable fact is that there is also a microscopic expression for
the same escape-rate in terms of the dynamical properties of the
moving particle in the scattering region. To understand the
microscopic formula we need to realize that although the moving
particle will eventually escape the region with probability $1$, there
are an infinite number of possible trajectories for the particle
whereby it never leaves the scattering region, i.e., it is never
absorbed at the walls. The set of initial positions and velocities of
the moving particle, that is, the set of initial points in the phase
space of the moving particle, that lead to orbits which are entirely
within the scattering region is called the repeller for this system,
and we denote it by ${\cal R}$. The repeller is typically a fractal
set of Lebesgue measure zero in the phase space, but with a
non-countable number of points. ${\cal R}$ will have a Hausdorff dimension
which is slightly less than the dimension of the phase space. The
escape-rate $\gamma$ can be expressed in terms of the dynamical
properties of the trajectories that are entirely confined to the
repeller as
\begin{equation}
\gamma = \sum_{i}\lambda_{+,i}({\cal R}) - h_{KS}({\cal R}),
\label{30}
\end{equation}
where the summation is over all the positive Lyapunov exponents,
$\lambda_{+,i}({\cal R})$ for
trajectories on the repeller, and $h_{KS}({\cal R})$ is the KS entropy
for these trajectories \cite{ott,gasni,gasdo,tel}. Note that we have a system for which Pesin's
theorem does not hold, but by a small amount of order $L^{-2}$. To get
some feeling for the origin of the escape-rate formula, though by no
means a derivation, we return to our argument for the validity of
Pesin's theorem \cite{jrdhvb}. We want to find the rate at which we are obtaining
information about the location of the initial points on the
repeller. As before, the stretching mechanism provides an exponential
rate of information growth about the initial location of the point in
phase space, but the probability that the system has not disappeared
through the boundary is decreasing exponentially, too. Therefore, we
should expect that the rate of information growth for points on the
repeller is obtained by combining these two exponential rates as
\begin{equation}
e^{t h_{KS}(\cal R)}=e^{[t\sum_{i}\lambda_{+,i}({\cal R})]}e^{-\gamma
t},
\label{31}
\end{equation}
or, equivalently, Eq. (\ref{30}) above. Simple one dimensional maps
which illustrate the derivation of the escape-rate formula, and
exhibit the structure of the fractal repeller are described in
Refs. \cite{ott,jrd,jrdhvb,gasholian}. As an example the reader might consider the map
\begin{equation}
\begin{array}{ccccc}
x' & = & 3x \,\,\, &  {\rm for} &  \,\,\, 0 \leq x \leq 1/2 \\
   & = & 3x-2 \,\,\, &  {\rm for} & \,\,\, 1/2 < x \leq 1
\end{array}
\label{32}
\end{equation}
and suppose that points mapped outside the interval $[0,1]$ have
escaped. For this map the fractal repeller is the middle third Cantor
set, the escape rate is $\gamma = \ln(3/2)$, the Lyapunov exponent on
the repeller is $\ln 3$, and the KS entropy is $\ln 2$. These results
can be understood by realizing that the map stretches all intervals by
a factor of $3$, accounting for the Lyapunov exponent, and that
dynamics on the repeller can be coded by sequences to two symbols,
each symbol appearing with equal probability, accounting for the KS
entropy on the repeller.

If we now combine the macroscopic expression for the escape-rate, Eq. (\ref{29}) with
the microscopic one, Eq. (\ref{30}), we obtain an expression for the
coefficient of diffusion, $D$, a transport coefficient, in terms of the
dynamical properties of the repeller,
\begin{equation}
D = \lim_{L \rightarrow \infty}\frac{L^2}{a}\left[
\sum_{i}\lambda_{+,i}({\cal R}) -h_{KS}({\cal R})\right],
\label{33}
\end{equation}
where we have taken the thermodynamic limit to ensure a result that
does not contain finite size effects. Eq. (\ref{33}) is due to Gaspard
and Nicolis \cite{gasni}. Gaspard and Dorfman have extended this method to apply to
the other transport coefficients, such as the shear and bulk viscosities,
and thermal conductivity of a fluid, as well as for chemical reaction
rates \cite{dogas}. When applying the escape-rate formalism to systems like the
Lorentz gas, for example, one usually finds that the Lyapunov
exponents and KS entropy for the repeller are equal to their infinite
system values plus very small corrections of order $L^{-2}$. Thus the
transport coefficients are contained in these small corrections to the
infinite system values, and the transport process is ``coded" in the
fractal 
structure of the repeller. 

The escape-rate method has been applied to compute the properties of
the fractal repeller for a number of model systems, including the
multibaker model for diffusion in one dimensional systems \cite{gasjsp}, the
two dimensional periodic Lorentz gas at sufficiently high density that
the free path length of the moving particle is small compared to the
dimensions of the system \cite{gasholian,gasbar}, to the random Lorentz gas at low
densities \cite{vbdprl1,lvbdprl3}, but still with a mean free path small compared to the size of
the system, and to Lorentz lattice gases \cite{ednj}. In these cases the Lyapunov exponents and KS entropies on
the repeller differ from their values in the thermodynamic limit, by
terms of order $L^{-2}$. The chief difficulty in using the escape-rate
method as a method to compute transport coefficients is that it is as
hard or harder to compute the dynamical quantities, particularly
$h_{KS}({\cal R})$ as it is to compute the transport coefficient directly, using
kinetic theory or other methods. It would be extremely valuable,
for example, to have analytic methods to compute $h_{KS}({\cal R})$ directly
rather than having to compute $\gamma$ and $\lambda_{+,i}({\cal
R})$ and using the escape rate to obtain the KS entropy on the
repeller, as one has to do in the analytic studies of the random
Lorentz gas.  

\subsection{The Gaussian Thermostat Method}

Another way to relate transport coefficients to dynamical quantities,
in this case Lyapunov exponents, was developed as a result of efforts
to simulate transport processes on the computer using molecular
dynamics techniques. It was found early on that the simulations of
viscous flow were plagued by the viscous heating of the fluid. To deal
with this Evans, Hoover, Nos\'e, and co-workers developed a
thermostatting method that keeps either the kinetic or total energy
constant in the fluid undergoing shear \cite{thermos}. Although the thermostat
destroys the symplectic, Hamiltonian structure of the dynamics, we get
as compensation, so to speak, a useful and interesting connection
between the transport coefficients in the thermostatted system and the
Lyapunov exponents for the thermostatted dynamical system, assuming
the system approaches a non-equilibrium steady state.  The ``destruction"
of the symplectic structure is
something of an overstatement. Liverani and Wojtkowski \cite{livwojt2} have shown that
these thermostatted systems are conformally symplectic, with useful
consequences for the Lyapunov spectrum, and Dettmann and Morriss \cite{dettmo2} have
shown that the thermostatted dynamics can be mapped onto Hamiltonian
dynamics with unusual canonical position and momentum variables. We
refer to their papers for details.

 We illustrate the
method again for a Lorentz gas with a  particle
moving among hard disk or sphere scatterers \cite{jrd,jrdhvb,morron,cels}. We suppose that the
scatterers are arranged randomly in space or in a regular
configuration with a finite free path for the moving particle. We also
provide the moving particle with a charge $q$ and suppose that it is
acted upon by an external electric field $\vec{E}$, in addition to the
scatterers. If there were no thermostat, then the average kinetic energy of
the moving particle would increase steadily with time, since the
collisions do not affect the kinetic energy, but the field does. To
counter this average increase in kinetic energy, we introduce a
``thermostat" which keeps the kinetic energy of the moving particle
constant between collisions with the scatterers. The equation for the
motion of the particle between collisions is given by
\begin{eqnarray}
\dot{\vec{r}} & = & \vec{p} \nonumber \\ 
\dot{\vec{p}} & = & \vec{E}-\alpha \vec{p},
\label{34}
\end{eqnarray}
where we have set both the mass, m, and the charge, q, of the moving particle
equal to unity. Here $\alpha$ is a function of the momentum and
electric field that is fixed by the condition that the kinetic energy
be constant, i.e. $\vec{p}\cdot \dot{\vec{p}}=0$. Thus
\begin{equation}
\alpha = \frac{\vec{E} \cdot \vec{p}}{p^2}.
\label{35}
\end{equation}
These equations of motion are supplemented by the equations for elastic
collisions of the particle with the scatterers,
\begin{equation}
\vec{p}\,' = \vec{p} -2({\hat{n}}\cdot \vec{p})\hat{n},
\label{36}
\end{equation}
where $\vec{p}\,'$ is the momentum of the particle after collision, and
$\hat{n}$ is a unit vector in the direction from the center
 of the scatterer to the point of contact with the moving particle at
the instant of collision. These equations of motion define a
reversible (under $\vec{p}\rightarrow  -\vec{p}$, 
$\vec{r}\rightarrow \vec{r}$, and $t \rightarrow -t$), non-Hamiltonian system in which the
phase space volume is not conserved. This latter remark follows from
the observation that
\begin{equation}
\frac{\partial \dot{\vec{r}}}{\partial \vec{r}} +\frac{\partial
\dot{\vec{p}}}{\partial \vec{p}} = -(d-1)\alpha,
\label{37}
\end{equation}
where $d$ is the number of spatial dimensions of the system. 

These equations have a number of remarkable consequences. Computer
simulations show that the system develops a steady state, and that in this steady state
the average value of $\alpha$ is positive, $\left<\alpha \right>_{ss}
>0$, as expected if the thermostat is to keep the system at constant
kinetic energy \cite{morron,vbdcpd,delpo1}. Now if the phase space volume is not conserved, but
decreases on the average, but the system still settles into a steady
state, at least within the resolution of the computer experiment, then it may be heading to a fractal attractor, of lower
dimension than the phase space dimension, here, $2d-1$. We will give a
simple example of a system that does just this in a moment. Moreover
one may define a Gibbs entropy, $S_{G}$ for this system by
\begin{equation}
S_{G}=-k_{B}\int \int
d{\vec{r}}d{\vec{p}}f(\vec{r},\vec{p},t)\left[\ln
f(\vec{r},\vec{p},t)-1\right],
\label{38}
\end{equation}
where we have assumed the existence of a phase space distribution
function, $f(\vec{r},\vec{p},t)$, for the moving particle. If this
function exists and is differentiable, then it satisfies the
conservation equation
\begin{equation}
\frac{\partial f}{\partial t}+\nabla_{\vec{r}}\cdot
(f\dot{\vec{r}})+\nabla_{\vec{p}}\cdot(f\dot{\vec{p}}) = 0.
\label{39}
\end{equation}
We now use this conservation equation to compute the time rate of
change of the Gibbs entropy due to the thermostat\cite{cels}. Here we
suppose that the Gibbs entropy of the system does not change with time
if the thermostat were not present, as is true for systems whose
distribution functions satisfy the usual Liouville equation. Now by assuming that the distribution function
vanishes at the boundaries (in space and momentum) of the system, we find that
\begin{equation}
\frac{d S_{G}(t)}{dt} = -k_{B}\left< \alpha \right>(d-1),
\label{40}
\end{equation}
where the angular brackets denote an average with respect to the phase
space distribution function $f$. This result that the entropy
decreases with time can be understood, to some extent, by noting that
if the system is heading toward an attractor it is getting localized into a more
restricted region of phase space with a concomitant decrease of the
Gibbs entropy. The conservation equation, Eq. (\ref{39}), has yet
another consequence of some considerable interest, namely, we
express $f$ as $f(\vec{r},\vec{p},t) = 1/V(\vec{r},\vec{p},t)$, where
$V$ is the volume of a small phase space region about
$(\vec{r},\vec{p})$ where the system is located at time $t$. Then the
equation for $f$, $\dot{f}=(d-1)\alpha f$, easily obtained from
Eqs. (\ref{37}, {39}), leads to
\begin{eqnarray}
\frac{d \ln V(t)}{dt} & = & -(d-1)\alpha \,\,\,{\rm or} \nonumber \\
\left<\frac{d \ln V(t)}{dt}\right> & = & -(d-1)<\alpha>.
\label{41}
\end{eqnarray}
Since the rate of change of small volume elements in phase space is
governed by the sum of the Lyapunov exponents,
\begin{equation}
V(\vec{r},\vec{p},t) \sim \exp \left[
t\sum_{i}\lambda_{i}(\vec{r},\vec{p})\right],
\label{42}
\end{equation}
where we may imagine that the Lyapunov exponents depend upon the phase
space point, and the sum is over {\it all} of the exponents, not just
the positive ones. From this it follows that
\begin{equation}
(d-1)\left< \alpha \right> = -\left<
\sum_{i}\lambda_{i}(\vec{r},\vec{p})\right>.
\label{43}
\end{equation}
Therefore, we
 can relate the average friction coefficient to the average value of
the sum of the Lyapunov exponents \cite{thermos}. 

We now push this analysis into the realm of transport theory by
returning to the calculation of the rate of change of the Gibbs
entropy. In a non-equilibrium steady state, the rate of decrease of
the Gibbs entropy must be at least matched by an increase in entropy
of the heat reservoir that is removing the heat generated by the
dynamical processes in the system. If we identify this positive rate of
entropy production with the irreversible entropy production required
by irreversible thermodynamics $\sigma E^{2}/k_{B}T$, where $\sigma$ is the
coefficient of electrical conductivity, $T$ is the thermodynamic
temperature, and $k_{B}$ is Boltzmann's constant, we find that
\begin{equation}
\sigma = -\frac{k_{B}T}{E^2}\left< \sum_{i}\lambda_{i}\right>.
\label{44}
\end{equation}
Eq. (\ref{44}) is the main result of this analysis \cite{thermos}. It shows that a
transport coefficient, in this case the electrical conductivity, is
proportional to the average rate of phase space contraction as
measured by the sum of all of the Lyapunov exponents. We will comment
in the final section upon the obviously problematic issues concerning
the use of entropy production arguments here.  

For small fields the electrical conductivity of the moving particle
and its diffusion coefficient in the Lorentz gas are directly related
by a simple factor, $D=[(p^2/(q^{2}m)] \sigma$,
and Eq. (\ref{44}) then provides another expression for the diffusion
coefficient in terms of dynamical quantities, the sum of all of the
Lyapunov exponents of the system. Similar expressions can be obtained
for other transport coefficients. The coefficient of shear viscosity,
in particular has been studied in great detail for particles that
interact with short ranged repulsive potentials, i.e. WCA particles  \cite{thermos,ecm}. 

The evaluation of these expressions for transport coefficients
appears to be formidable since one has to determine the complete
Lyapunov spectrum for the thermostatted system in an external
field. This situation has been simplified to a great extent by the
observation, and in some cases the proof \cite{livwojt2,ecm,dettmo}, of a {\it conjugate pairing
rule} for these thermostatted systems. That is, the non-zero Lyapunov exponents
can be ordered in conjugate pairs, $\lambda_{i},\lambda_{-i}$, say,
such that the sum of each conjugate pair is the same for all conjugate
pairs. Consequently, the sum of all of the Lyapunov exponents is equal
to the product of the number of conjugate pairs and the sum of one
conjugate pair. This is of course a generalization of the conjugate
pairing rule for symplectic systems, where the sum is zero. The
conjugate pairing rule for thermostatted systems has been observed in
computer simulations, and has been proved analytically for a number of
systems, provided the thermostat keeps the total kinetic energy \cite{livwojt2,dettmo},
rather than the total energy, kinetic plus potential, constant. Of course for hard sphere type
systems, there is no difference between the kinetic and total
energies.

\subsection{A Simple Map with an Attractor}
 
In order to see how a fractal attractor might form in the
phase space for a thermostatted system we turn once again to a simple
two dimensional map. Here we construct a map that has some of the features that we expect to
see in a thermostatted Lorentz gas, for example. That is, we consider
a map with reversible, hyperbolic dynamics, but allow for a phase space volume
contraction in part of the phase space. Such a map is provided by a
map $\Phi$ of the unit square onto itself given by
\begin{equation}
\begin{array}{ccccc}
\Phi (x,y) & = & (x/l, ry) \,\,\, & {\rm for} &  \,\,\, 0\leq x \leq l \\  
& = & \left( (x-l)/r, r +ly) \right) \,\,\, & {\rm for} & \,\,\, l<x \leq 1,
\end{array}
\label{45}
\end{equation}
where $l,r > 0, l \neq r$, and $l+r=1$. This map is illustrated in
Fig. 8. One can easily calculate the positive and negative Lyapunov
exponents for this map and they are
\begin{eqnarray}
\lambda_{+} & = & l\ln \frac{1}{l} + r \ln \frac{1}{r} , \\
\lambda_{-} & = & l \ln r + r \ln l, \,\,\, {\rm and} \\
\lambda_{+} + \lambda_{-} & = & (l-r)\ln \frac{r}{l} \leq 0.
\end{eqnarray}
Here the equality holds in the last equation only if $l=r=1/2$. 
The SRB theorem tells us that the time average of any continuous
function defined on the unit square will approach an ensemble average
taken with respect to the appropriate SRB measure. Some reflection will
convince the reader that the SRB measure is smooth in the expanding, or $x$, direction
but fractal in the contracting, or $y$, direction. The details of this and related SRB
measures are described in Ref. \cite{tasaki}. We say that the system has
evolved to a fractal attractor with a smooth measure in the $x$
direction and fractal in the $y$ direction. This attractor is
invariant under the map $\Phi$ since the limiting form of the set is
not changed by the map, or for that matter its inverse, $\Phi^{-1}$
which can easily be constructed. Another interesting point about
this map $\Phi$ is related to its reversibility property.         The
map $\Phi^{-1}$ has exactly the same Lyapunov
exponents as the map $\Phi$. However the SRB measure for the inverse
map is smooth in the expanding $y$ direction but fractal in the
contracting $x$ direction, and the time reversed attractor is also
invariant under the two maps, $\Phi$ and its inverse. Sometimes we
call this pair of invariant sets an attractor with a corresponding
repeller. Depending on the direction of the time, one of the two invariant sets is an
attractor, and the other is a repeller.

\subsection{Fractal Forms for Diffusion Coefficients}

Our discussions in this section have been largely formal. To
conclude this section, we will consider the case of deterministic
diffusion in one dimension under the action of a simple, piecewise
linear map, such as that illustrated in Fig. 9. This map has the
general form $x_{n+1}=M(x_n)$, where $M(x)$ is a piecewise linear
function of $x$, and satisfies the condition $M(x+1)=1+M(x)$. In the
interval $0 \leq x \leq 1$, $M(x)$ is given explicitly by
\begin{equation}
\begin{array}{ccccc}
M(x) & = & ax \,\,\, & {\rm for} & \,\,\, 0 \leq x \leq 1/2 \\
     & = & ax +(1-a) \,\,\, & {\rm for} & \,\,\, 1/2 < x \leq 1, 
\end{array}
\label{49}
\end{equation}
where the slope of the map $a$ satisfies the condition $a >2$, which
is necessary for diffusion to take place at all. R. Klages studied
this and similar maps for his Ph. D. work \cite{rkth}. The diffusion coefficient
could be obtained by mapping this system onto a Markov process using a
one dimensional version of Markov partitions. All of the appropriate
details are given in Klages' dissertation and in Refs. \cite{rkjrd1,rkjrd2}. Here we
will consider the shape of the diffusion coefficient, $D$, as a function of
the slope $a$ for $a >2$. The values of $D$ for even integer $a$ can
easily be found by mapping this process onto a simple random walk on a
line with
a non zero probability, depending on $a$, of staying at the same
site. For odd integer $a$, the problem is not much harder and $D$ can
also be obtained easily. The method of Markov partitions gives $D$ for
a dense set of values of $a$ on the real line, and this is used to
give the general structure of the dependence of $D$ upon $a$. This
dependence is illustrated in Fig. 10. There we see that $D$ is a
fractal function of $a$ with quite a rich fractal structure. This
rather surprising structure shows that transport coefficients are not
always simple functions of the parameters of the system. Further
studies on models of this type are underway by Klages \cite{Klages} and
by J. Groeneveld \cite{groen}.

We have
established some interesting relations between dynamical and transport
quantities and showed that fractal structures in the microscopic phase space have
properties that are directly related to macroscopic transport
properties of the system. In the next section we look at questions of
the dynamical foundations of irreversible
kinetic equations and show how such equations can be used to compute,
among other things, Lyapunov exponents and KS entropies, for the
Lorentz gas, as an example.

\section{The Dynamical Foundations of Kinetic Equations}

Two ideas have been developed in the previous sections which seem to
be central to an understanding of the dynamical origins of
irreversible phenomena in fluids. In particular, we can now isolate
some of the key features that must be taken into account when one
tries to understand the efficacy of stochastic equations such as the
Boltzmann transport equation for describing irreversible processes in
gases. Briefly stated, these two central ideas are:

 (1) One can, for
Anosov systems and probably with suitable modifications for hard sphere systems as well, separate
the tangent space at a particular point in phase space and, in
fact, phase space itself, into expanding, contracting and center manifolds. The expanding and contracting manifolds exchange behavior
on the time reversed motion. 

(2) Measures tend to be smooth in the
unstable, or expanding, directions, and fractal in the stable,
or contracting, directions. 

Therefore one can imagine the following scenario for the
derivation of an irreversible equation from a dynamical
analysis. Consider some initial ensemble distribution in phase space
that is a measureable function of the phase space variables with
respect to the microcanonical measure, say. To be specific consider an
initial distribution which is unity on a certain set $A$ of small
measure and zero elsewhere. In the course of time, this set will
be stretched along the unstable directions, becoming more and more
uniformly distributed in these directions, but forming more and more
of a fractal structure in the stable directions. If then, we project
the distribution function onto the unstable directions, say by
integrating over the stable directions and thereby obtaining a reduced
distribution function, we ought to see the approach to a uniform
distribution function in the projected variables. 

To make this argument less abstract and more transparent, we consider
its application to
our familiar models, the baker map
\cite{jrd,berry,reichl} and the Arnold cat map. To begin we
will need the version of Liouville's equation appropriate for maps
that that evolve at discrete time intervals, called the
Perron-Frobenius equation \cite{ott,lasmak}. 

\subsection{The Perron-Frobenius Equation}

We consider a map of some region $R$ onto itself. To be definite we
take $R$ to be two dimensional, and the map to be of the form
\begin{equation}
(x',y') = \Phi(x,y)=(\Phi_{1}(x,y), \Phi_{2}(x,y)).
\label{50}
\end{equation}
We consider $R$ to be our phase space, and we suppose there is some
phase space distribution evolving from some initial distribution on
$R$ denoted by $\rho_{0}(x,y)$. Then after $n$ iterations of the map,
a new distribution function, $\rho_{n}(x,y)$, is obtained which satisfies
the recursion equation
\begin{eqnarray}
\rho_{n}(x,y) = \int_{R} \int dx' dy'
\delta(x-\Phi_{1}(x', y'))\delta(y-\Phi_{2}(x', y')) \cdot \nonumber \\
\rho_{n-1}(x', y'). 
\label{51}
\end{eqnarray}
This recursion equation, the Perron-Frobenius equation, is self-evident provided the delta functions
are not to be evaluated at possible discontinuity points of the
distribution function. The
delta functions can be evaluated in terms of the pre-image points of
$(x,y)$, and the Jacobian of the map $\Phi$ at these pre-image points.
The examples to which we apply this equation are simple area preserving maps on the
unit square for which the Jacobian of $\Phi$ is simply unity, but more general
cases can be treated as well.

\subsection{The Baker's Map}

To apply the Perron-Frobenius equation to the baker's map, we take $R$
to be the unit square, and the mapping $\Phi$ to be given by
Eq. (\ref{5}). An elementary calculation leads to the recursion
equation
\begin{equation}
\begin{array}{ccccc}
\rho_{n}(x, y) & = & \rho_{n-1}(x/2, 2y)\,\,\, & {\rm for} & \,\,\, 0
\leq y <1/2 \\
& = & \rho_{n-1}(\frac{x+1}{2}, 2y-1) \,\,\, & {\rm for} & \,\,\, 1/2 \leq
y \leq 1.
\end{array}
\label{52}
\end{equation}

For this map the unstable direction is the $x$-direction, and we can
define a reduced distribution function that only depends on $x$ by
integrating over the $y$ variation of $\rho_{n}(x,y)$. We therefore
define $W_{n}(x)$ by
\begin{eqnarray}
W_{n}(x) & = & \int_0^1 dy\rho_{n}(x,y) \nonumber \\
& = & \int_0^{1/2} dy \rho_{n-1}(x/2, 2y) + \int_{1/2}^1 dy
\rho_{n-1}(\frac{x+1}{2}, 2y-1) \nonumber \\
& = & \frac{1}{2} \left[
W_{n-1}(\frac{x}{2})+W_{n-1}(\frac{x+1}{2})\right]
\label{53}
\end{eqnarray}
This equation can be considered to be a model ``kinetic" equation
since it is obtained from the discrete form of Liouville's equation by
integrating over the $y$ variation. The reader may also notice that
Eq. (\ref{53}) is itself the Perron-Frobenius equation for the map $x'
= 2x$ mod $1$. It is easy to see that $W_n =1$,
or any non-zero constant, is a solution of this equation, and we
require that $\rho_{n}$ and $W_{n}$ be non-negative so that they can
represent probability distributions. Moreover there is an $H$-theorem
associated with Eq. (\ref{53}), very similar to Boltzmann's
$H$-theorem. That is, we define $H_n=\int_0^1 dx W_n(x)\ln W_n(x)$,
and elementary inequalities based upon the observation that a straight
line
connecting two points on the curve $z = u \ln u$ lies below the curve
show that $H_n \leq H_{n-1}$. The equality sign holds only if $W_{n}$
is a constant. Finally, Eq. (\ref{53}) can be solved if $W_0(x)$ is a
measurable function of $x$, and any arbitrary, normalized  initial
distribution will approach $1$ as $n \rightarrow \infty$. It is
important to note that we have done nothing more than integrate the
Perron-Frobenius equation for this map over the coordinate in the stable direction,
assuming only that the initial distribution function is well
behaved. As the time increases, the reduced
distribution function becomes an increasingly smoother function of $x$, finally reaching an
equilibrium distribution, in a monotonic, irreversible manner. Now if
we were to start with the same initial distribution function $\rho_{0}(x,y)$, but
follow the time reversed map, then the unstable direction is parallel
to the $y$ axis, and the reduced distribution function would be obtained by
integrating the Perron-Frobenius equation for the time reversed motion
over $x$. The resulting equation
 would be identical to Eq. (\ref{53}), with $x$ replaced by $y$. The
reversibility of the underlying equations of motion is still preserved
but irreversible equations are obtained by selecting the direction of
time and projecting the full distribution function onto the unstable
manifold. The reader is also invited to see what happens if the
distribution function is projected onto the stable manifold, for
the baker's map. The paper by Tasaki and Gaspard \cite{tasgas} should prove useful
for this analysis.  

\subsection{The Arnold Cat Map}

As an extension of the ideas just discussed we consider now the Arnold
cat map. The analysis of the corresponding Perron-Frobenius equation
is complicated by the fact that the unstable and stable directions are
not oriented along the $x$ and $y$ axes respectively, but are rotated
by a fixed angle. Instead of working out the mathematics let us just
look at the map on the computer. We will be interested in the reduced
distributions projected onto the $x$-axis or onto the $y$-axis. This
is a little closer to what we need to understand the Boltzmann
equation since for a dilute gas with $N$ particles we do not know the directions of the stable
and unstable manifolds in phase space, but we do consider a projected
or reduced distribution function, namely the single particle
distribution function obtained by integrating the Liouville
distribution over the phases of all but one particle.

We start with the same initial distribution function for the Arnold
cat map as illustrated in
Fig. 4, that is, where all of the distribution is concentrated in the
lower left corner of the unit square. In Fig. 11 we plot the
distribution function projected onto the $x$-axis, for various
times starting with a step function at $t=0$. Note that this function
also becomes smooth with time and appears to have reached equilibrium
after four steps. Now consider the same map but project the
distribution function onto the $y$-axis. this is illustrated in
Fig. 12. Exactly the same things happen. This reduced distribution
function also approaches an equilibrium value after a ten time steps
or so. One can also show that $H$-functions defined in terms of either
of these distribution functions decrease monotonically in time. If one
reverses the motion, then very similar things will happen: both
of the reduced distribution functions, whether projected on the $x$ or
$y$ axes will approach equilibrium. Only if we project somehow onto
the stable direction will we not see an evolution to a smooth equilibrium distribution.

We conclude from a study of these maps that the approach of a reduced distribution function to a smooth
equilibrium value depends only on the construction of a reduced
distribution function that is obtained by projecting the full
distribution function onto a direction that is not orthogonal to
unstable directions in phase space. Of course this is really a
conjecture. It is an important and open problem to work this out in
detail for more realistic systems. Furthermore, we might be asking too
much of realistic systems that this picture be realized in essence for
them. Instead only much weaker properties may really be necessary for
the variables of physical interest, such as the pressure exerted by a gas on the
walls of the container, to show an approach to equilibrium. However, for Anosov
systems, at least, where there is a very clear decomposition of the
entire phase space into stable, unstable, and center manifolds, the picture sketched above is likely to hold.

\subsection{The Random Lorentz Gas}

In this subsection we briefly indicate how the Boltzmann equation
itself may be used to calculate the Lyapunov exponents of a simple
model of a gas, the hard disk Lorentz model \cite{vbdprl1,lvbdprl3,vbdcpd}. To do this we combine the
ideas of Boltzmann with those of Ya. G. Sinai \cite{sinai}, thereby suggesting the
existence of a deep connection between ``molecular chaos''-the
stochastic hypothesis that underlies the Boltzmann transport equation- and
dynamical chaos, which shows that stochastic-like behavior is possible
even for deterministic mechanical systems.

Consider a collection of hard disks of radius $a$ placed at random on
a plane with number density $n$, such that $na^{2} <<1$. The moving
particle always travels with the same speed, $v$ and makes elastic
collisions with the scatterers. To calculate
the Lyapunov exponent we consider two trajectories on the constant
energy surface that are infinitesimally close to each other, but
separating, nonetheless. We consider the separation of a diverging
bundle of trajectories in position space, since if this separation grows
exponentially, so will the separation in velocity space, since both
have to grow with the same exponential. We suupose without loss of
generality that if we follow the trajectory
bundle backward in time all the trajectories go through a
common point of intersection. The separation of two trajectories in
the bundle can be
expressed in terms of a radius of curvature, $\rho$, which is the
distance from the present position of the trajectories to the point of
intersection. If we denote the distance between the two trajectories
by $\Delta$, one finds that (see Fig. 13) 
\begin{equation}
\frac{d \Delta}{d t}=\frac{v \Delta}{\rho}
\label{54}
\end{equation}
This equation has the simple solution
\begin{equation}
\Delta (t) =\exp \left( \int_0^t d \tau
\frac{v}{\rho(\tau)}\right)\Delta(0).
\label{55}
\end{equation}
It now follows immediately, from the fact that this system is ergodic that the positive Lyapunov exponent,
$\lambda_{+}$, is given as
\begin{equation}
\lambda_{+} = v \left< \frac{1}{\rho} \right>,
\label{56}
\end{equation}
where the brackets denote an ensemble average. Therefore we need to
determine the ensemble distribution for the radius of curvature. This
is the point where the Boltzmann equation becomes useful. Let us
consider an ``extended" distribution function, $F(\vec{r},\vec{v},\rho,t)$, which includes the
position of the moving particle, $\vec{r}$, its velocity, $\vec{v}$ the
time, $t$, and
the radius of curvature, $\rho$, as variables. In a sense this is a two
particle distribution function since the separation of two
trajectories is included as one of the variables. Now we know how
$\vec{r}$ and $\vec{v}$ change during free motion and at a collision
with a scatterer. The radius of curvature $\rho$ grows linearly with
time during the free motion as $\rho (t + dt)=\rho (t) +v\,dt$, and
changes from a precollision value $\rho^{-}$ to a post collision value
$\rho^{+}$ at a collision, where 
\begin{equation}
\frac{1}{\rho^{+}} = \frac{1}{\rho^{-}} + \frac{2}{a \cos \phi},
\label{57}
\end{equation}
where $\phi$ is the angle of incidence at collision\cite{sinai}. Typically
$\rho^{-}$ is on the order of the mean free path between collisions,
and $\rho^{+}$ is always less than or equal to $a/2$. 

Using standard methods of kinetic theory one can derive the following
equation for $F(\vec{r},\vec{v},\rho,t)$, the extended Lorentz-Boltzmann
equation
\begin{eqnarray}
\left[ \frac{\partial}{\partial t}+\vec{v} \cdot \nabla +
v \frac{\partial}{\partial\rho} \right] F(\vec{r},\vec{v},\rho,t) = - \nu
F(\vec{r},\vec{v},\rho,t)    \nonumber \\
+(\nu/2)\int_{-\pi/2}^{\pi/2} d \phi \int_{0}^{\infty} d \rho ' \cos
\phi \delta \left( \rho -\frac{(a\cos\phi)/2}{1 + (a\cos \phi)/(2\rho
')} \right) \nonumber \\
\times F(\vec{r},\vec{v}', \rho ',t),
\label{58}
\end{eqnarray}
and $\vec{v}' = \vec{v} -2(\hat{n}\cdot \vec{v})\hat{n}$
, as in Eq. (\ref{36}), and $\nu = 2 nav$ is the average collision frequency
for the moving particle. This equation can be solved easily for an
equilibrium system where $F$ does not depend on $\vec{r}$ or $t$. The
details of the derivation and the solution are given elsewhere\cite{jrdhvb}, but
here we give the final result
\begin{equation}
\lambda^{+} = 2nav(1 - C - \ln(2na^2)),
\label{59}
\end{equation}
where $C$ is Euler's constant. We emphasize the point that not only
does dynamical systems theory allow us to understand, to some extent
at least, the dynamical origins of kinetic theory, kinetic theory in
turn can be used to compute important quantities needed to show that
the ideas  we have been discussing are indeed relevant for our
understanding of irreversible processes in fluids.

\section{Conclusions}

Our excursion into the dynamical foundations of the kinetic theory of
gases has provided us with a very stimulating and suggestive picture
of irreversible processes. If we can show that a typical system like a
gas of hard spheres or WCA particles is indeed an Anosov system, then
many striking consequences are immediate. In particular, the phase
space has a tangent space almost everywhere which can be decomposed
into unstable, stable, and neutral subspaces. Measures and distribution
functions become smooth with time along the unstable directions,
approaching equilibrium values as the time gets very large compared to
characteristic microscopic times. The reversibility of the equations
of motion reflects itself in the interchange of stable and unstable
directions under time reversal, so that equilibrium states appear in
the time reversed motion as the distribution function becomes smooth
along the unstable directions of the time reversed motion. 
  
If this picture is correct then we can relate microscopic dynamical quantities
such as Lyapunov exponents and KS entropies to macroscopic quantities
such as transport coefficients, Here we illustrated this connection
using the escape-rate formalism, and the method of Gaussian
thermostats. However, these are only two of a number of useful methods
for making this connection. For example, methods based on periodic orbit
expansion or Ruelle-Pollicott resonances are under very active study
at the present time
\cite{gasbook,gaspre}.
 
We conclude with a few remarks:

1) We have presented a picture here of how things might be for a
typical system of interest to those applying kinetic theory methods to
compute transport properties. It is not known whether this picture is
indeed correct. We must study a number of systems of increasingly more
realistic structures in order to get deeper insights into this
question. Progress is being made in this direction using both computer
and analytical studies of a number of different systems. 

2) We know very little about the complicated fractal structures that
underlie transport processes in fluids, i.e., the attractors and
repellers that we have discussed earlier. Only for simple two
dimensional maps do we have any good intuition about these
fractals. Even higher dimensional maps would provide useful examples
for more complicated systems.

3) We have not mentioned many aspects of dynamical systems theory
that are relevant for a further analysis of the topics discussed
here. In particular, the thermodynamic formalism of Sinai, Bowen and
Ruelle \cite{ruelle,becksch} provides a close, but often very formal, connection between
dynamical systems theory and equilibrium statistical mechanics. This
connection is surely not accidental, and can be very helpful for
understanding the dynamical foundations of statistical mechanics, in
general.

4) The reader may have been surprised by the appearance of a negative
entropy production in the discussion of the Gaussian thermostatted
systems in Sec. 3.2. This negative sign is really not mysterious,
since we are gaining more information about the system as its phase
space trajectory becomes more and more localized on an attractor. In equilibrium thermodynamics one can reduce the
entropy of a gas by applying work in a reversible, isothermal compression. However, the entropy
of the universe stays constant when one takes into account the entropy change of
the thermostat that works on the gas. Here, similarly, one
has to consider the system {\it plus} the thermostat in order to make
sense of the entropy production. In a steady state the total entropy
production must be zero, so the negative entropy production of the
system must be matched by a positive entropy production in the
thermostat.

A number of papers have been written recently which discuss this topic
in more detail. Ruelle has proved that the entropy production by the
thermostat must be strictly positive \cite{ruellejsp}. Tel, Vollmar and
Breymann have discussed the entropy production for a simple model
based upon the baker map from the point of view
of coarse graining the phase space \cite{tvb,mr}. Here the point is that
we are unable, experimentally, to follow the localization of the
thermostattes system onto the attractor because we are limited to the
resolution of our observing devices. Therefore beyond a certain point
we will not see the entropy of the thermostatted system
decrease. Instead we will see the entropy of the coarse grained,
observed, system increase as the microscopic processes take place on a
scale that is too fine for our measuring devices. In the Tel, Vollmer,
and Breymann analysis the positive entropy production is taken to be
the difference between the information available in a coarse grained description where nothing is
changing beyond a certain scale and the information available in a
complete microscopic description where one can follow the trajectory
onto the attractor. There still needs to be a study where one connects
this idea of coarse graining to a physical description of a thermostat
and shows that including the entropy production in the thermostat is
equivalent to coarse graining the phase space of the thermostatted system.

5) P. Gaspard has discussed entropy production for a simple
reversible system which consists of a chain of baker's maps coupled in
such a way that a density gradient may be established and
maintained. He shows that the positive entropy production associated
with this process is connected to the fractal structure that develops
in the system as the steady density gradient is established. The phase
space density is not a differentiable function of the phase space
variables, and one can only define a coarse grained entropy for the
system. Gaspard shows that this coarse grained entropy is positive an
its production has the form required by irreversible thermodynamics.

6) We have not mentioned quantum mechanics at all. It seems
inconsistent with our physical understanding of matter to restrict our
attention exclusively to classical systems. The quantum versions of the
ideas discussed here are still in the early phases of development and
we refer the reader to the literature for further details
\cite{casati}. This is an
area where even our understanding of even the simplest systems is not
entirely secure.

To conclude, there is a lot of work still to do in order to undertand
in detail the chaotic foundations of transport theory and to provide a
clear microscopic explanation of all of the phenomena that we associate with
irreversible processes in  fluids.

ACKNOWLEDGEMENTS: The author would like to thank Henk van Beijeren,
E. G. D. Cohen, Matthieu Ernst, Pierre Gaspard, Edward Ott, and Tamas
Tel for many helpful conversations as well as Thomas Gilbert for many
helpful discussions on SRB measures. He would like to thank Charles
Ferguson and Rainer Klages for supplying a number of the figures, and
he would also like to thank them as well as Thomas Gilbert, Luis Nasser, and Debabrata Panja for
many helpful and clarifying discussions, and especially Ramses van Zon for a critical reading of
this manuscript. He thanks the  National Science Foundation for support
under Grant PHY-96-00428.

\newpage

\centerline{\bf FIGURE CAPTIONS}

{\it Figure 1}. The tent map on the unit interval $0 \leq x \leq 1$.
\vskip 0.5 cm
{\it Figure 2}. The baker's map on the unit square $0 \leq x,y \leq 1$.
\vskip 0.5 cm
{\it Figure 3}. The Arnold cat map on the unit square.
\vskip 0.5 cm
{\it Figure 4}. The initial set $A$ is confined to the lower left
corner of the unit square. The dynamics of the points in this set is
governed by the Arnold cat map.
\vskip 0.5 cm
{\it Figure 5}. The evolution of the set $A$ after $2$ iterations of
the Arnold cat map.
\vskip 0.5 cm
{\it Figure 6}. The evolution of the set $A$ after $3$ iterations of
the map.
\vskip 0.5 cm
{\it Figure 7}. The evolution of the set $A$ after $10$
iterations. The initial set $A$ consisted of $10^{5}$ points, and the
continuous nature of the initial set is no longer preserved. With more
points, this evolved set would appear to be a set of closely spaced
parallel lines, nearly convering the unit square uniformly.
\vskip 0.5 cm
{\it Figure 8}. A simple map on the unit square with an attractor. The
map is given by Eq. (45), and the areas of regions I and II
individually are not preserved by the map.
\vskip 0.5 cm
{\it Figure 9}. The one-dimensional piecewise linear map given by
Eq. (49). Here $a > 2$.
\vskip 0.5 cm
{\it Figure 10}. The diffusion coefficient $D$ as a function of the
slope of the map $a$, for the map in Figure 9. Curve (a) is the
function over the range $2 \leq a \leq 8$. Curves (b)-(f) show the
diffusion coefficient at various magnifications. The error bars are
too small to be noticeable on these graphs.
\vskip 0.5 cm
{\it Figure 11}. The $x$-distribution function for the Arnold cat map
at iterations $n = 0,1,2,3,4$, showing that the distribution becomes
nearly uniform after $4$ iterations. The initial condition is the same
as in Figure 4. These curves are affected by a small computer
generated error. 
\vskip 0.5 cm
{\it Figure 12}. The $y$-distribution function for the Arnold cat map
at the same iterations as in Figure 11, for the same initial
conditions.
\vskip 0.5 cm
{\it Figure 13}. The growth of the radius of curvature between
collisions.
\vskip 0.5 cm 
{\it Figure 14}. The change in the radius of curvature at collision.


\begin{thebibliography}{99}

\bibitem{ott}E. Ott, {\it Chaos in Dynamical Systems}, Cambridge
University Press, Cambridge, (1992).

\bibitem{poincare}H. Poincar\'e, {\it New Methods in Celestial
Mechanics}, D. L. Goroff, Ed., A. I. P. Press, New York, (1993).

\bibitem{lebpen} J. Lebowitz and O. Penrose, Physics Today {\bf
26}, 23, (1973).

\bibitem{rueck}J.-P. Eckmann, and D. Ruelle, Rev. Mod. Phys. {\bf 57},
617, (1985).

\bibitem{gasbook}P. Gaspard, {\it Chaos, Scattering, and Statistical
Mechanics}, Cambridge University Press, Cambridge, (to appear).

\bibitem{guckho}J.Guckenheimer and P. Holmes, {\it Nonlinear
Oscillations, Dynamical Systems, and Bifurcations of Vector Fields},
Springer Verlag, Berlin, (1983).

\bibitem{yorkebo} K. T. Alligood, T. D. Sauer, and J. A. Yorke, {\it
Chaos: An Introduction to Chaotic Systems}, Springer Verlag, new York, (1997).

\bibitem{jrd}J. R. Dorfman, {\it From Molecular Chaos to Dynamical
Chaos}, Lecture Notes, University of Utrecht, and University of Maryland, College Park, (1997).

\bibitem{jrdhvb}J. R. Dorfman and H. van Beijeren, Physica A{\bf 240},
12, (1997).

\bibitem{dohvb}J. R. Dorfman and H. van Beijeren, in B. Berne (Ed.),
{\it Statistical Mechanics, B}, Plenum Press, New York, (1977).

\bibitem{resdel} P. Resibois and M. deLeener, {\it Classical Kinetic
Theory of Fluids}, Wiley, New York, (1977).

\bibitem{chapcow}S. Chapman and T. G. Cowling, {\it The Mathematical
Theory of Non-Uniform Gases}, 3rd. Ed., Cambridge University Press,
Cambridge, (1970).

\bibitem{bowen}R. Bowen, {\it Equilibrium States and the Ergodic
Theory of Anosov Diffeomorphisms}, Lecture Notes in Mathematics,
No. 470, Springer Verlag, Berlin, (1975); {\it On Axiom A
Diffeomorphisms}, Regional Conference Series in Mathematics, No. 35,
American Mathematical society, Providence, (1978); R. Bowen and
D. Ruelle, Invent. Math. {\bf 29}, 181, (1975). 

\bibitem{ruelle}D. Ruelle, {\it Thermodynamic Formalism},
Addison-Wesley, Reading Ma., (1978); Am. J. Math., {\bf 98}, 619,
(1976); Phys. Math. IHES, {\bf 50}, 275, (1979); {\it Elements of
Differentiable Dynamics and Bifurcation Theory}, Academic Press, New
York, (1989); D. Ruelle and
Ya. G. Sinai, Physica A{\bf 140}, 1, (1986). 

\bibitem{sinai}Ya. G. Sinai, {\it Introduction to Ergodic Theory},
Princeton University Press, Princeton, (1976); Russ. Math. Surv. {\bf
25}, 137, (1970); {\it Dynamical Systems}, World Scientific,
Singapore, (1991); Russ. Math. Surv., {\bf 21}, 21, (1972); {\it
Topics in Ergodic Theory}, Princeton University Press, Princeton, (1994). 

\bibitem{kathas}A. Katok and B. Hasselblatt, {\it Introduction to the
Modern Theory of Dynamical Systems}, Cambridge University Press,
Cambridge, (1995).

\bibitem{gasni}P. Gaspard and G. Nicolis, Phys. Rev. Lett., {\bf 65},
1693, (1990).

\bibitem{thermos} D. J. Evans and G. P. Morriss, {\it Statistical
Mechanics of Nonequilibrium Liquids}, Academic Press, London, (1990);
W. G. Hoover, {\it Computational Statistical Mechanics}, Elsevier,
Amsterdam, (1991). See also papers collected in M. Mareschal and
B. Holian, (Eds.), {\it Microscopic Simulations of Complex
Hydrodynamic Phenomena}, Plenum Press, New York, (1992). 

\bibitem{uhlfo} G. E. Uhlenbeck and G. W. Ford, {\it Lectures in
Statistical Mechanics}, American Mathematical Society, Providence, (1963).

\bibitem{arnav}V. I. Arnold and A. Avez, {\it Ergodic Problems of
Classical Mechanics}, Benjamin, New York, (1968).

\bibitem{birkh}G. D. Birkhoff, Proc. Nat. Acad. Sci. {\bf 17}, 656, (1931).

\bibitem{szasz}D. Szasz, Stud. Scient. Math. Hungarica, {\bf 31}, 299,
(1996); Physica A{\bf 194}, 86, (1993).

\bibitem{szsim} N. Simanyi and D. Szasz, ``Ergodicity of Hard Spheres
in a Box", (to be published).

\bibitem{livwojt} C. Liverani and M. Wojtkowski, ``Ergodicity in
Hamiltonian Systems" in {\it Dynamics Reported (New Series)}, Vol
4. Springer Verlag, Berlin (1995), p. 130.

\bibitem{gallor}G. Gallavotti and D. Ornstein, Comm. Math. Phys., {\bf
38}, 83, (1974).

\bibitem{sinai2}Ya. G. Sinai, Funct. Anal. Appl. {\bf 13}, 192, (1980).

\bibitem{peterson}K. Peterson, {\it Ergodic Theory}, Cambridge
University Press, Cambridge, (1983).

\bibitem{billings}P. Billingsley, {\it Ergodic Theory and Information},
Wiley, New York, (1965).

\bibitem{arnold} V. I. Arnold, {\it Mathematical Methods in Classical
Mechanics}, 2nd. Ed., Springer Verlag, Berlin, (1989).

\bibitem{lasmak}A. Lasota and M. C. Mackey, {\it Chaos, Fractals, and
Noise} 2nd. ed. Springer Verlag, Berlin, (1994).

\bibitem{pesin}Ya. Pesin, Russ. Math. Surv. {\bf 32}, 55, (1997).

\bibitem{gasdo}P. Gaspard and J. R. Dorfman, Phys. Rev. E{\bf 52},
3525, (1995).

\bibitem{lanford}O. E. Lanford in {\it Chaotic Behavior of
Deterministic Systems}, G. Iooss, R. H. G. Helleman and R. Stora,
Eds., North Holland, Amsterdam, (1983), p. 6.  

\bibitem{galco}G. Gallavotti and E. G. D. Cohen, J. Stat. Phys. {\bf
80}, 931, (1995).

\bibitem{tel}T. Tel in {\it Directions in Chaos}, Vol. 3, Hao Bai-Lin,
Ed., World Scientific, Singapore, (1990). 

\bibitem{gasholian}P. Gaspard and F. Baras, in  M. Mareschal and
B. Holian, (Eds.), {\it Microscopic Simulations of Complex
Hydrodynamic Phenomena}, Plenum Press, New York, (1992), p. 301. 

\bibitem{dogas}J. R. Dorfman and P. Gaspard, Phys. Rev. E{\bf 51}, 28,
(1995).

\bibitem{gasjsp} P. Gaspard, J. Stat. Phys. {\bf 68}, 673, (1992).

\bibitem{gasbar}P. Gaspard and F. Baras, Phys. Rev. E{\bf 51}, 5332, (1995).

\bibitem{vbdprl1}H. van Beijeren and J. R. Dorfman,
Phys. Rev. Lett.,{\bf 74}, 4412, (1995); {\bf 76}, 3238(E), (1996).

\bibitem{lvbdprl3}A. Latz, H. van Beijeren, and J. R. Dorfman, 
Phys. Rev. Lett. {\bf 78}, 207, (1997).

\bibitem{ednj}M. H. Ernst, J. R. Dorfman, R. Nix, and D. Jacobs,
Phys. Rev. Lett. {\bf 74}, 4416, (1995).

\bibitem{livwojt2} M. P. Wojtkowski and C. Liverani, ``Conformally
Symplectic Dynamics and Symmetry of the Lyapunov Spectrum", preprint, (1997)

\bibitem{dettmo2}C. P. Dettmann and G. P. Morriss, Phys. Rev. E{\bf
54}, 2495, (1996); {\bf 55}, 3693, (1997).

\bibitem{morron}G. P. Morriss, C. P. Dettmann, and L. Rondoni, Physica
A{\bf 240}, 84, (1997); J. P. Lloyd, M. Niemeyer, L. Rondoni, and
G. P. Morriss, Chaos, {\bf 5}, 536, (1995).

\bibitem{cels}N. I. Chernov, G. I. Eyink, J. L. Lebowitz, and
Ya. G. Sinai, Comm. Math. Phys. {\bf 154}, 569, (1993).

\bibitem{vbdcpd}H. van Beijeren, J. R. Dorfman, E. G. D. Cohen,
Ch. Dellago, and H. A. Posch, Phys. Rev. Lett. {\bf 77}, 1974, (1996). 

\bibitem{delpo1} Ch. Dellago, H. A. Posch, and W. G. Hoover,
Phys. Rev. E{\bf 53}, 1485, (1996).

\bibitem{ecm}D. J. Evans, E. G. D. Cohen, and G. P. Morriss, Phys
Rev. A{\bf 42}, 5990, (1990).

\bibitem{dettmo}C. P. Dettmann and G. P. Morriss, Phys. Rev. E{\bf 53},
R5541, (1996).


\bibitem{tasaki}S. Tasaki (private communication and to be published).

\bibitem{rkth}R. Klages {\it Deterministic Diffusion in
One-Dimensional Maps} Ph. D. Dissertation, Technical University of
Berlin, Berlin, (1995).

\bibitem{rkjrd1}R. Klages and J. R. Dorfman, Phys. Rev. Lett. {\bf
74}, 387, (1995).

\bibitem{rkjrd2}R. Klages and J. R. Dorfman, Phys. Rev. E{\bf 55},
R1247, (1997). 

\bibitem{Klages}R. Klages (unpublished)

\bibitem{groen}J. Groenveld (unpublished)

\bibitem{berry}M. V. Berry, in S. Jorna, (Ed.), {\it Topics in
Nonlinear Dynamics}, A. I. P. conference Proceedings, No. 46, American
Institute of Physics, New York, (1978).

\bibitem{reichl}L. Reichl, {\it A Modern Course in Statistical
Mechanics}, University of Texas Press, Austin, (1980).

\bibitem{tasgas} S. Tasaki and P. Gaspard, J. Stat. Phys. {\bf 81},
935, (1995).

\bibitem{gaspre} P. Gaspard, Phys. Rev. E{\bf 53}, 4379, (1996). 

\bibitem{becksch} C. Beck and F. Schl\"ogl, {\it Thermodynamics of
Chaotic Systems}, Cambridge University Press, Cambridge, (1993).

\bibitem{ruellejsp} D. Ruelle, J. Stat Phys. {\bf 85}, 1, (1996).

\bibitem{tvb}T. Tel, J. Vollmer, and W. Breymann, Europhys. Lett. {\bf
35}, 659, (1996); Phys. Rev. Lett. {\bf 77}, 2945, (1996).

\bibitem{mr} G. P. Morriss and L. Rondoni, Physica A{\bf 233}, 767, (1996).

\bibitem{gasent}P. Gaspard, Physica A{\bf 240}, 54, (1997);
J. Stat. Phys. (to appear).

\bibitem{casati} G. Casati and B. Chirikov (Eds.), {\it Quantum Chaos},
Cambridge University Press, Cambridge, (1995); K. Nakamura, {\it
Quantum Chaos}, Cambridge University Press, Cambridge, (1994); see
also CHAOS, {\bf 3}, No. 4, (1993).

\end{thebibliography}
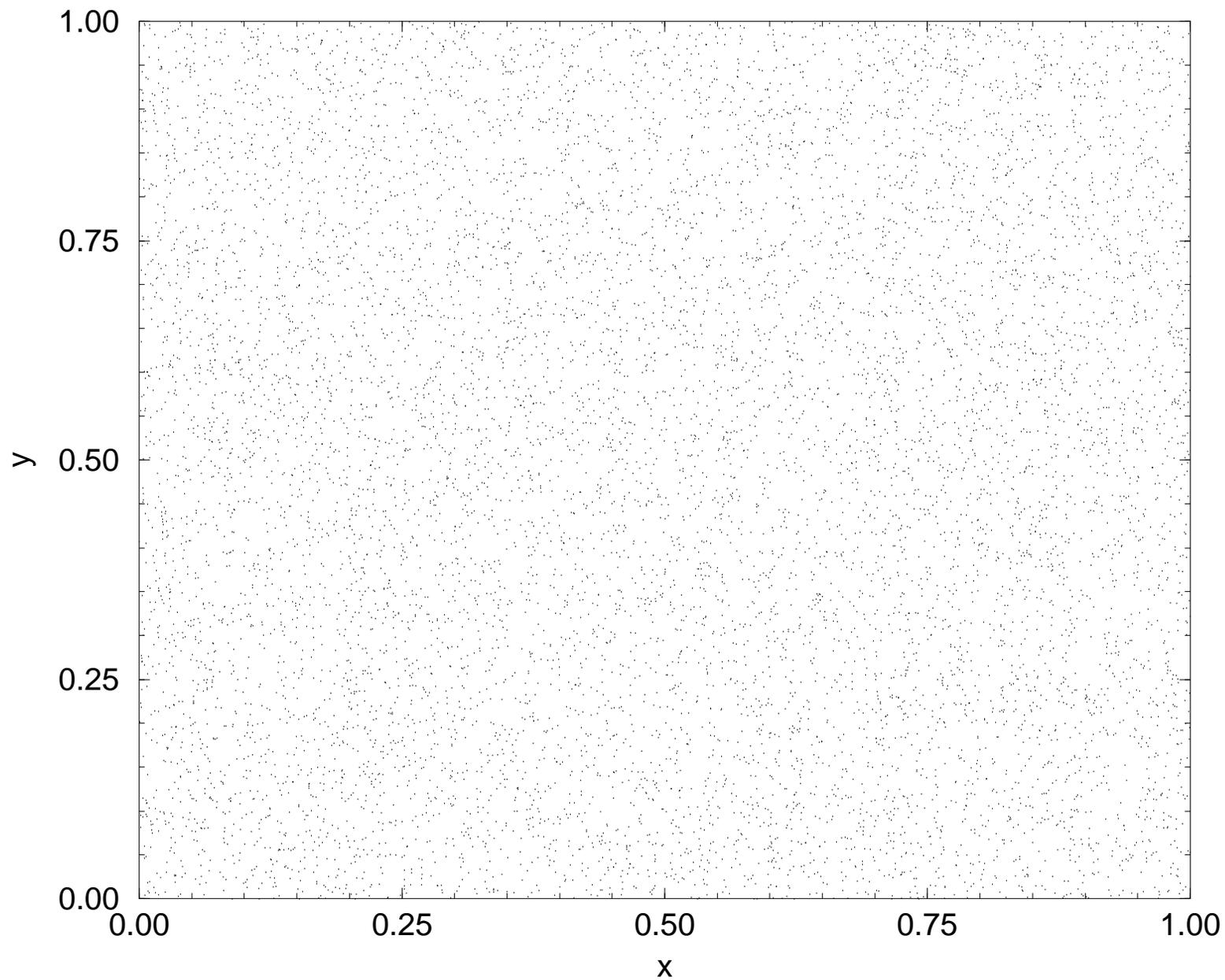
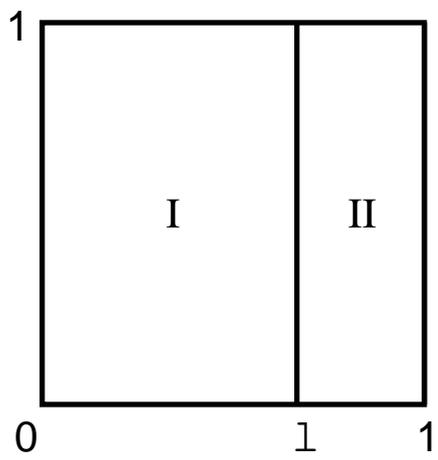
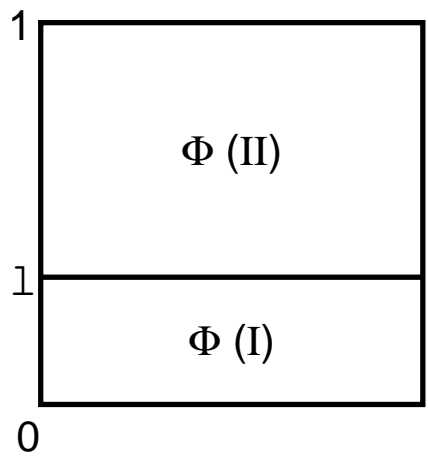
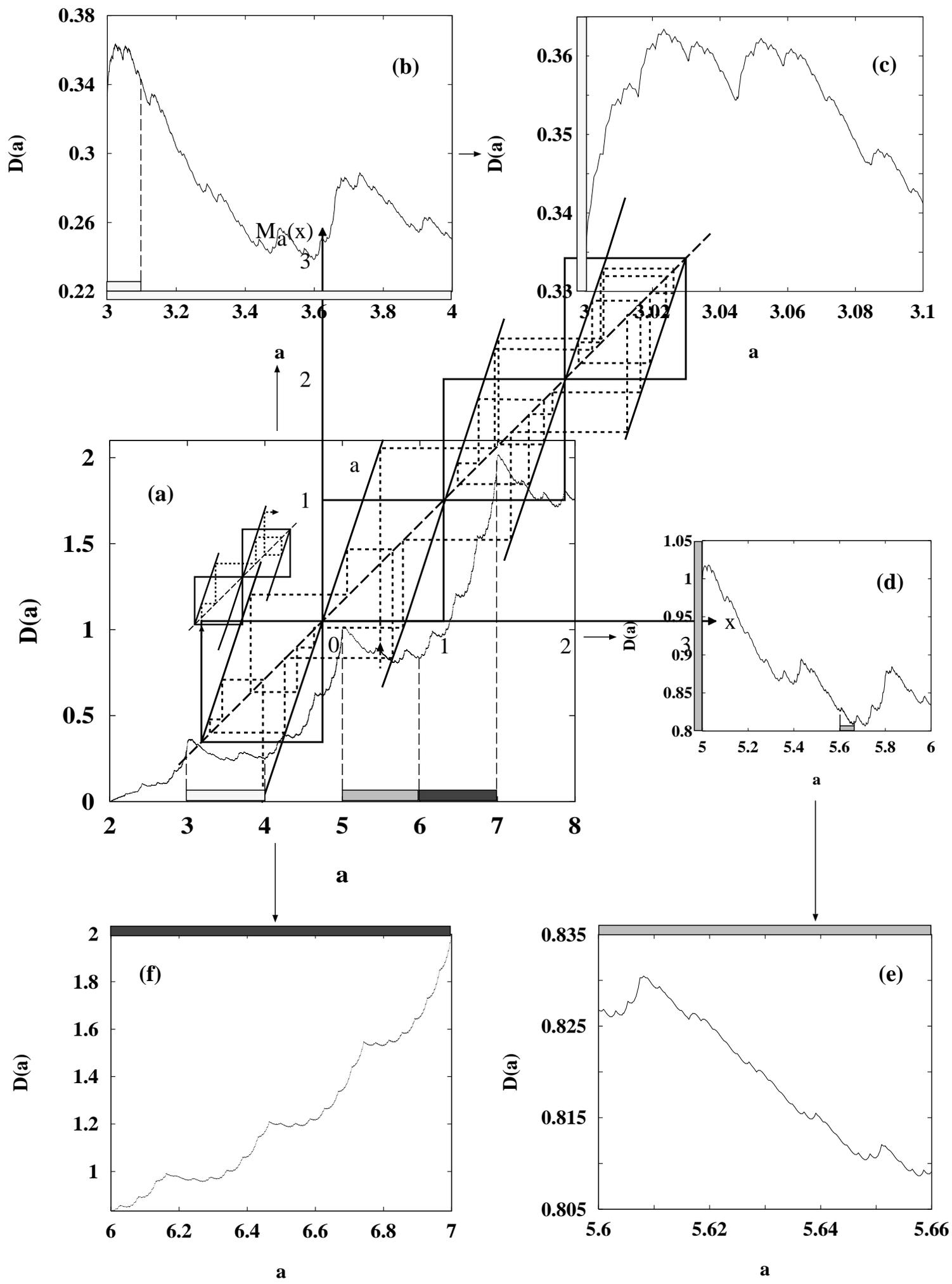
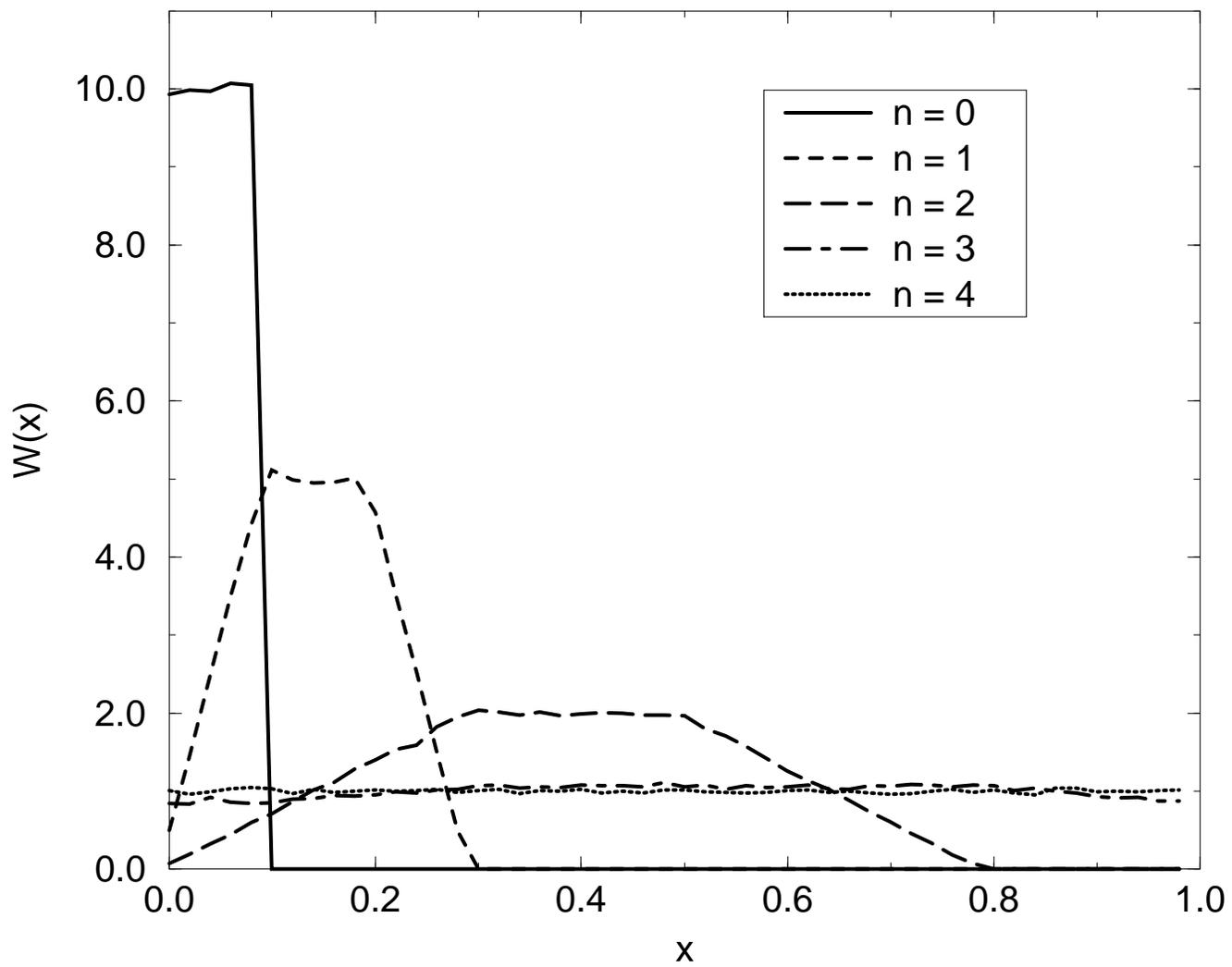
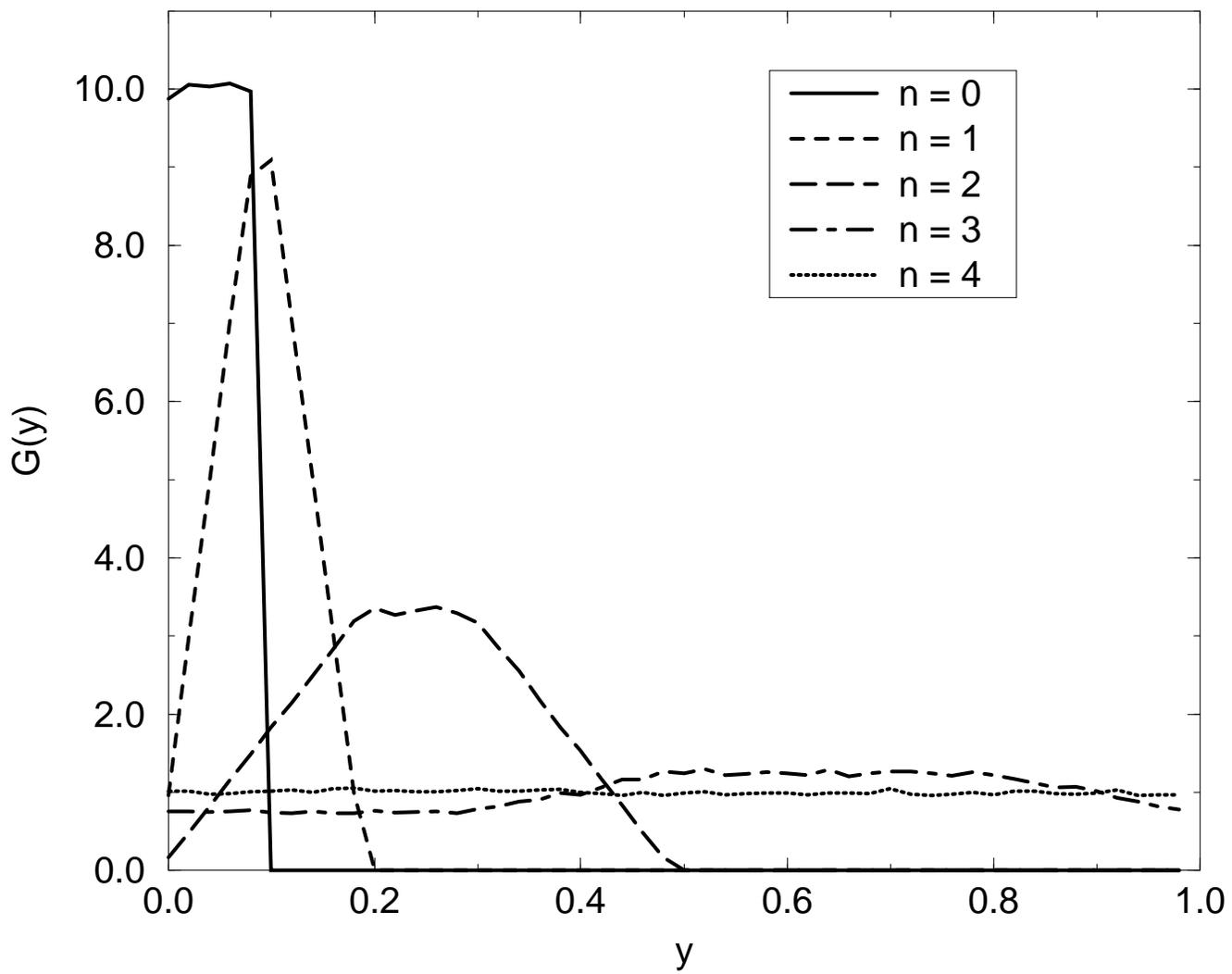
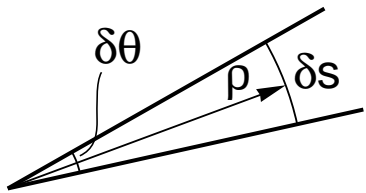
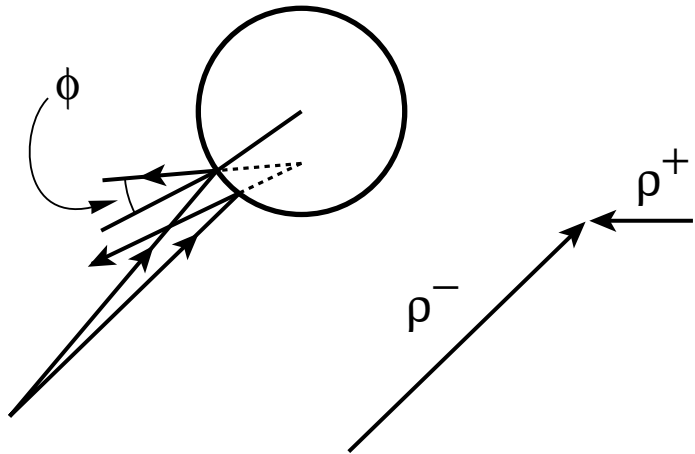
\end{document}